\documentclass[12pt,prd, aps, showpacs, superscriptaddress,floatfix]{revtex4-2}
\usepackage{graphicx}
\usepackage{bm}
\usepackage{multirow}
\usepackage{epsfig}
\usepackage{psfrag}
\usepackage{soul}
\usepackage{multirow}
\usepackage{mathptmx}
\usepackage{mathrsfs}
\usepackage{amsmath, amssymb}
\usepackage{dcolumn}
\usepackage{epstopdf}
\usepackage{color}
\usepackage[T1]{fontenc} 
\usepackage{hhline}
\usepackage{changebar}
\usepackage{placeins}

\usepackage{slashed}

\usepackage{flexisym,breqn,gensymb}
\usepackage{subcaption}
\usepackage[colorlinks=true, linkcolor=blue, filecolor=blue, urlcolor=blue, citecolor=blue,plainpages=false]{hyperref}

\newcommand\cu{Physics Department, Columbia University, New York,
  NY 10027, USA}
\newcommand\bnl{Brookhaven National Laboratory, Upton, NY 11973, USA}

\makeatletter
\newsavebox{\@brx}
\newcommand{\llangle}[1][]{\savebox{\@brx}{\(\m@th{#1\langle}\)}%
  \mathopen{\copy\@brx\kern-0.7\wd\@brx\usebox{\@brx}}}
\newcommand{\rrangle}[1][]{\savebox{\@brx}{\(\m@th{#1\rangle}\)}%
  \mathclose{\copy\@brx\kern-0.7\wd\@brx\usebox{\@brx}}}
\makeatother

\newcommand{\bbar}[1]{\overline{#1}}
\newcommand{\trueavg}[1]{\llangle #1 \rrangle}
\newcommand{\ensavg}[1]{\langle #1 \rangle}
\newcommand{\bigtrueavg}[1]{\llangle[\Bigg] #1 \rrangle[\Bigg]}
\newcommand{\bigensavg}[1]{\Bigg\langle #1 \Bigg\rangle}
\newcommand{\calC}{{\cal C}}

\begin{document}

\title{Bootstrap-determined p-values in Lattice QCD}

\author{N.H.~Christ}\affiliation{\cu}
\author{R.~Eranki}\affiliation{\cu}
\author{C.~Kelly}\affiliation{\bnl}

\collaboration{RBC and UKQCD Collaborations}

\date{September 17, 2024}

\maketitle
\centerline{ABSTRACT}

We present a general method to determine the probability that stochastic Monte Carlo data, in particular those generated in a lattice QCD calculation, would have been obtained were that data drawn from the distribution predicted by a given theoretical hypothesis. Such a probability, or p-value, is often used as an important heuristic measure of the validity of that hypothesis. The proposed method offers the benefit that it remains usable in cases where the standard Hotelling $T^2$ methods based on the conventional $\chi^2$ statistic do not apply, such as for uncorrelated fits. Specifically, we analyze a general alternative to the correlated $\chi^2$ statistic referred to as $q^2$, and show how to use the bootstrap as a data-driven method to determine the expected distribution of $q^2$ for a given hypothesis with minimal assumptions.  This distribution can then be used to determine the p-value for a fit to the data. We also describe a bootstrap approach for quantifying the impact upon this p-value of estimating population parameters from a single ensemble of $N$ samples. The overall method is accurate up to a $1/N$ bias which we do not attempt to quantify.  

\tableofcontents

\section{Introduction}

Lattice field theory has become a valuable tool to determine the predictions of a continuum field theory when perturbation theory cannot be applied.  However, as an approach that relies on statistical methods to evaluate the Feynman path integral describing the field theory of interest, this method requires a careful interpretation of statistical errors.  In this paper we will focus on a widely used application of lattice field theory methods, lattice QCD; the non-perturbative study of quantum chromodynamics.  

In many cases, the physical quantities (particle masses, decay couplings, etc) are computed by fitting one or more Green's functions obtained from a lattice calculation to a theoretical fit ansatz. Such an analysis of lattice data will be imperfect, because the theoretical fit function perforce neglects certain physical phenomena present in the lattice data, for example the presence of excited states. A disagreement between the theoretical fit function and the lattice data will also arise because of systematic errors in the lattice calculation itself not described by the theoretical fit function, such as the effects of the finite lattice spacing or the finite volume. Both of these difficulties lead to errors in the final results which may be obscured by the statistical fluctuations of the lattice data.  Thus, it is important to understand these statistical fluctuations and to distinguish whether the difference between the best fit and the data is caused by these fluctuations or by an actual failure of the fit ansatz to describe the data.  Developing an improved method  to characterize such a ``goodness-of-fit'' is the goal of this paper.

The conventional strategy for assessing the goodness-of-fit via frequentist statistical methods involves computing a ``p-value" representing the probability that the data agrees with the fit ansatz up to only statistical errors. To compute this quantity one typically relies upon analytic descriptions of the fluctuations of the data, usually the $\chi^2$ or Hotelling $T^2$ distributions~\cite{hotelling1931}.  These descriptions are based on certain assumptions that may or may not be reliable for a lattice simulation, namely that either the number of samples is very large ($\chi^2$) or that the underlying data are normally distributed ($T^2$). Both descriptions also assume that the samples are statistically independent, and that fully correlated fits are being employed with the covariance matrix computed in the conventional manner. A typical lattice calculation with a few hundred samples whose underlying probability distribution is the QCD path integral, and which are generated via Markov-chain Monte Carlo methods and thus potentially possess subtle long-range autocorrelations, may violate one or more of these assumptions, thus rendering the goodness-of-fit metric unreliable. This is particularly true for cases where low statistics necessitates employing approximate descriptions of the covariance matrix, e.g. uncorrelated fits or truncated eigenspectrum methods, which completely invalidate the analytic descriptions, often resulting in the abandonment of any attempt to estimate the goodness-of-fit.

In this document we introduce a data-driven approach for estimating the p-value based on the bootstrap method. This strategy requires minimal assumptions and can be applied to arbitrary manipulations of the covariance matrix and in the presence of autocorrelations.  A similar objective is addressed in Ref.~\cite{Bruno:2022mfy} for the specific case of normally distributed data and a certain class of modifications to the standard definition of $\chi^2$.   We believe that in some cases our approach can be powerful, allowing one to exploit a customized definition of the covariance matrix designed to reduce the statistical errors in the resulting fit parameters while retaining the ability to assess the goodness-of-fit.

This paper is organized as follows: In Section~\ref{sec-overview} we will provide a comprehensive overview of the conventional and proposed bootstrap approaches to allow the reader to more easily follow the underlying philosophy of the approach. In Section~\ref{sec-conventional-approach} we provide more detail on the conventional approach and introduce many important concepts, and in Section~\ref{sec-toy-model-demos} we demonstrate these concepts with toy data. In Section~\ref{sec-bootstrap-null-dist} we detail the bootstrap approach and provide toy data examples for statistically-independent data samples. Finally, in Section~\ref{sec-autocorrelated-data} we discuss how this approach can be extended to account for autocorrelations in the lattice data, and also discuss a particular modification of the covariance matrix strategy that has been used to successfully control the statistical errors on the RBC \& UKQCD collaboration's calculation of the $I=0$ $K\to\pi\pi$ decay amplitude~\cite{Kelly:2019wfj,RBC:2020kdj}, for which the bootstrap approach allowed the estimation of the goodness-of-fit under circumstances where conventional approaches could not be applied.

\section{Overview}
\label{sec-overview}

To more clearly introduce the proposed methods, we will present a brief overview of the current approaches used to control such statistical effects and then summarize the additional technique proposed in this paper.  

The most common method used to test for agreement between stochastic lattice ``data'' and a particular fit function used to model that data is based upon the ``$\chi^2$" or ``loss" of the fit. For reasons that will become clear below, we will refer to this quantity instead as $q^2$ (cf. Eqs.~\eqref{eq-avg}, \eqref{eq-q2-def} and \eqref{eq-covmatdef}). The approach can be phrased in the language of hypothesis testing: We define the ``null hypothesis" as the hypothesis that the imperfections in the theoretical fit function and the lattice data described above are unimportant. If the data that we obtained are consistent with the null hypothesis ({\it i.e.} they are described by the fit function up to purely statistical fluctuations), the ensemble can be considered as a collection of random samples drawn from a hypothetical population whose mean is described precisely by our fit function for some (unknown) set of parameters. We will describe such ensembles as ``null-consistent". The distribution of $q^2$ obtained by fitting many such ensembles of the same size drawn from this population is referred to as the ``null distribution".

The ratio that gives the conventional definition of $q^2$ is indeed remarkable. In the limit of a large number of samples, $N\to\infty$, its null distribution is known analytically: it is the $\chi^2$ distribution, which depends only on the number of degrees of freedom (the number of data points being fit minus the number of fit parameters) and does not depend on the nature of the statistical fluctuations or of the fit function. The fact that the null distribution is known in advance allows us to perform a hypothesis test: we can compare the value of $q^2$ obtained by fitting to our data with the $\chi^2$ distribution to test whether those data are indeed consistent with the null hypothesis. This is achieved by computing a "p-value" -- essentially the probability of obtaining a null-consistent ensemble of data that would have a worse fit -- that can be treated as an indicator of the likelihood that the imperfections in our fit function or lattice data are indeed negligible.  

Unfortunately, the large number of samples needed to ensure that the $\chi^2$ distribution is the correct description of the null distribution are often impossible to obtain, and an understanding of the nature of the fluctuations in the statistical data becomes important.   As $N$ becomes smaller, often the first difficulty that arises in the analysis of the null distribution of $q^2$ is the additional fluctuation arising from the fact that the covariance matrix ${\cal C}(t,t')$, defined in Eq.~\eqref{eq-covmatdef}, is computed from the same, finite, sample.  However, with possibly reasonable assumptions, this case can also be described analytically: If we assume that the data are normally distributed, then the fluctuations in ${\cal C}(t,t')$ are described by the Wishart distribution, and the null distribution is described by the Hotelling $T^2$ distribution~\cite{hotelling1931}.  Only the assumption that the underlying lattice data follows a normal distribution, the number of data points being fit, and the number of fit parameters, are needed to determine the Hotelling $T^2$ distribution, so again we can use it for a hypothesis test by computing the p-value for our value of $q^2$ under this distribution.

In a lattice calculation, the underlying gauge configurations are distributed according to the QCD path integral and not a normal distribution. However, many quantities of interest derived from these configurations are obtained as averages over a number of independent and/or semi-independent measurements, for example by averaging correlation functions over propagator source locations or over multiple consecutive configurations (i.e., binning), and so, by the central limit theorem, may be sufficiently close to normal that the Hotelling description is reliable. Often, however, applying the full correlated fit procedure results in unacceptably-large statistical errors on the fit parameters due to large fluctuations in a poorly-resolved covariance matrix. In such cases, the typical approach is to replace the covariance matrix by an approximation capturing some portion of the correlation between the data points. 

Another challenge is the presence of autocorrelations in data samples generated by a Markov chain Monte Carlo process. These effects can be suppressed by a combination of binning and measuring only upon well-separated configurations in the Markov chain, but the latter involves discarding expensive and potentially-valuable data, and the former comes at the cost of reducing the number of (binned) samples from which the covariance matrix is obtained, thus enhancing its fluctuations and increasing the statistical error on the fit results. As a result, these procedures are typically only applied to a sufficient degree that the statistical error estimate stabilizes, rather than to completely eliminate long-range autocorrelations. This process also discards meaningful data that could be used to compute a custom covariance matrix that results in lattice QCD predictions with a reduced statistical error, as we describe in Sec.~\ref{sec-autocorrelated-data}.

Both residual autocorrelations and modifications to the covariance matrix invalidate the analytic descriptions of the null distribution and therefore make any quoted goodness-of-fit metrics based upon those analytic descriptions unreliable. In this document we introduce a strategy for obtaining an estimate of the null distribution directly from the data using the bootstrap method.  

The bootstrap method relies upon the similarity between the true distribution of a derived quantity, e.g. the ensemble mean, over many independent ensembles of data samples, each drawn at random from the underlying distribution, and an ``empirical distribution" obtained if that quantity is instead measured on a series of ensembles whose data samples are drawn with uniform weighting from those which appear in single random ensemble. Here we employ this method to obtain an empirical distribution of $q^2$. As the ensemble to which the bootstrap method is being applied grows in size, the bootstrap method will give an increasingly accurate $q^2$ distribution with a systematic bias that falls at least as fast as $1/N$.  While this should give confidence that the method is correct, the method is intended to be used when $N$ does not approach infinity, a limit in which distorted expressions for $q^2$ are not needed and conventional methods can be used.  Thus, an important part of this paper is the study of a series of relevant toy models in which the bootstrap results for the $q^2$ distribution can be compared to the ``true" distribution obtained by generating many independent ensembles (and, where applicable, the analytic distribution), and the identification of a window within which $N$ is sufficiently large that the bootstrap method is reliable but small enough that the usual $\chi^2$ or Hotelling $T^2$ descriptions cannot be correctly applied.

We will initially consider toy data drawn from a population with a known functional dependence, and use the same function to fit the data. In this case, the data are null-consistent by design, and the distribution of $q^2$ obtained by fitting many such ensembles defines the true null distribution. Taking one such ensemble as our proxy for the lattice data -- our ``original ensemble" -- we are interested in obtaining the corresponding p-value, which describes the probability of drawing a null-consistent ensemble whose $q^2$ value is equal or larger than the value we found. This probability can be estimated from the bootstrap empirical $q^2$ distribution obtained as we describe in the following paragraph.

By construction, the mean central value of the empirical distribution of a measurement obtained from some original ensemble is equal to the ensemble mean of that quantity.  However, due to statistical fluctuations, this mean value will not be perfectly described by our fit function. As a result, a single bootstrap ensemble drawn from that empirical distribution is not null-consistent even when the original ensemble itself is null-consistent. To correct for this, we "recenter" the empirical distribution by uniformly shifting each sample therein such that its means are described exactly by the fit function. As we do not (apart from in these toy examples) know the true population values for the parameters of this function, we substitute with those obtained by fitting our original ensemble, under the assumption that those values are sufficiently close to the population values that the effects on the $q^2$ distribution can be discounted. (We also provide a means to estimate this ``statistical error" in this document.) By performing this recentering, we explicitly enforce the null-consistency of the bootstrap data, while retaining the subtleties in the fluctuations of the data from the original ensemble. This bootstrap empirical null distribution can then be employed {\it in lieu} of a null distribution that is known {\it a priori} to compute the p-value for the fit to our original ensemble, that will be argued to be correct up to ${\cal O}(1/N)$ biases and the aforementioned statistical error.

The procedure described above provides a general means of obtaining the p-value for data that is assumed to be null-consistent. However, when performing a lattice calculation we are often more interested in using the p-value for a hypothesis test, to ascertain whether our data actually are null-consistent or otherwise.  A unrealistically small p-value, perhaps one significantly below 0.1, is typically interpreted as strong evidence that the data is not described by the fit function. 

In order to exploit our data-derived bootstrap approach in this context, it must continue to provide a good estimate of the null distribution even when applied to data that is {\it not null-consistent}. In order to achieve this we must first clarify the ambiguity as to how to {\it define} the corresponding null-consistent data: The null hypothesis described above specifies only the functional form of the population means of null-consistent data, but makes no assertions regarding the values of the parameters of that function, nor on how the samples fluctuate, the latter of which will dictate not just the distribution of ensemble means about the population mean but also how the estimate of the covariance matrix we compute from those samples will fluctuate between ensembles. With our toy models mentioned above, we made implicit choices for these components of our hypothesis when we specified the distributions from which we draw our data, and, apart from in the large $N$ limit or when the data are normally distributed, the null distribution will depend on these choices. We must therefore make our null hypothesis more explicit, identifying not only a functional form but a specific choice (or range of choices) for the parameters and a model for the fluctuations. The corresponding hypothesis test can then be used to obtain a likelihood that the lattice data are consistent with a population defined in this way.  Interpreted as such, a low p-value will indicate that one or more of these assertions is invalid. We repeat that the choice of the null hypothesis is entirely up to the scientist, and merely defines the question that is being asked of the data.

The above is made more challenging by the fact that, other than for normal data, the mean of an ensemble and its standard deviation -- and by extension the ensemble covariance matrix -- are not statistically independent. Therefore, we must actually model the combined probability distribution of the means and the ensemble covariance matrix, which enter the numerator and denominator of the expression for $q^2$, respectively. Fortunately, if we choose to base our model of the fluctuations on the those of our lattice data, the bootstrap method allows us to realize this choice. Recentering the resulting bootstrap ensembles then allows us to impose the specific functional form on the population means appropriate to our hypothesis in a way that does not affect the fluctuations of the samples around the ensemble mean or the fluctuations of the ensemble mean about the population mean. Note that here, the bootstrap resampling procedure is simply fulfilling the role of generating pseudo-independent ensembles, and were it replaced in this context by the generation of truly independent, non-null-consistent ensembles (as in the case for the toy models we will use to demonstrate this approach), a recentering would still be required to impose the functional form specific to our hypothesis.

The choice of parameters for our function is more straightforward: since we are primarily interested in testing the functional form and not some specific choice of parameters, we can simply choose those we obtain by fitting to our original data. In this way, if it happens that our original ensemble is indeed well described by our model, then the fit parameters should be close to the corresponding population parameters and we would be likely to obtain a good p-value, and thus pass our hypothesis test. 

We stress that the bootstrap method employed in this way can only test a very specific class of hypotheses: those for which the data are drawn from a population whose means are described by our chosen functional form with best-fit parameters and our data-derived model for the fluctuations of the samples. It cannot account for any changes in the distribution of the fluctuations that might occur if a different functional form than that tested truly describes the population. As a concrete example, consider a lattice simulation of a finite-volume system of two identical particles to which only a single energy eigenstate contributes. The true functional form describing the Euclidean-time dependence of this system would include a single exponential term for the two-particle state and a constant term describing the around-the-world temporal propagation of a single particle. Now imagine that we hypothesize that the around-the-world term can be ignored and thus fit with a single exponential form. The bootstrap procedure approximates the subtleties of the fluctuations in the population, including the effect of this around-the-world propagation, which would manifest as a large eigenmode of the covariance matrix, affecting its eigenspectrum and the condition number of the matrix. Thus, our model of the fluctuations maintains some information about the population that contradicts our model of the system's Euclidean-time dependence and that may also have a non-trivial impact on the distribution of $q^2$. Nevertheless, if the around-the-world term truly is significant, the fact that our functional form for the time dependence did not incorporate this term makes it highly unlikely that we would obtain a good p-value for our fit. As a result, the hypothesis test still remains a valid discriminator.

We conclude that this bootstrap-determined p-value has two important uses: First it is an accurate measure of the probability that we would have drawn an ensemble of samples with the given value of $q^2$ or larger, in the case that the population of data samples is described by our fit function with fluctuations captured through the bootstrap process {\it in lieu} of independent ensembles.  Second, it allows for a hypothesis test as to whether the data are likely to be drawn from a population with our chosen functional form and parameters obtained by fitting to our data, with the fluctuations again modeled on those of the data. This second, important use of the p-value is not compromised by its bootstrap origin.  This conclusion is also supported by the same bootstrap studies of the toy models referred to above.

\section{Conventional approach to fitting lattice data}
\label{sec-conventional-approach}

For the sake of simplicity, throughout this document we consider a single lattice correlation function $C_i(t)$ measured as a function of lattice time $t$ on each configuration $i\in\{1\ldots N\}$ of an ensemble. To avoid confusion, we will reserve the term ``sample" for describing a data point associated with a single configuration, and ``ensemble" as a collection of $N$ such points. In order to differentiate between various classes of average we will uniformly denote ensemble averages with either a bar, $\bbar{\phantom{C}}$ (for single quantities) or single angle brackets $\ensavg{\hphantom{.}}$  (for expressions), i.e.
\begin{dmath}
{ \bbar{C}(t) \equiv \ensavg{C(t)} \equiv \frac{1}{N}\sum_i C_i(t)\,. }
\label{eq-avg}
\end{dmath}
Averages over the population of independent ensembles will be denoted with a double angle-bracket, $\trueavg{\hphantom{.}}$. For example $\trueavg{\bbar{C}(t)}$ is the average over many ensembles of the ensemble mean of $C(t)$. As the ensemble mean is an unbiased estimator of the ``true" value of the underlying theory, the population average in this case corresponds to this true value.

We fit $C(t)$ to some function $f(t;\vec a)$ where $\vec a$ is the set of $A$ parameters that we wish to determine, by minimizing
\begin{dmath}
q^2 = \sum_{t,t'=t_{\rm min}}^{t_{\rm max}} \Big( \bbar{C}(t) - f(t;\vec a) \Big) {\cal C}^{-1}(t,t') \Big( \bbar{C}(t') - f(t';\vec a) \Big)\,, \label{eq-q2-def}
\end{dmath}
over variations of $\vec a$, where $t_{\rm min}$ and $t_{\rm max}$ describe the fit range. Here ${\cal C}$ is the covariance matrix, which parametrizes the fluctuations of $C(t)$ about the population mean,
\begin{dmath}
{\cal C}(t,t') \equiv \bigtrueavg{ \Big(\bbar{C}(t) - \trueavg{\bbar{C}(t)} \Big)\Big(\bbar{C}(t') - \trueavg{\bbar{C}(t')} \Big)   }
\approx \frac{1}{N-1}\bigensavg{ \Big( C(t) - \bbar{C}(t) \Big)\Big( C(t') - \bbar{C}(t') \Big)   }
\,.\label{eq-covmatdef}
\end{dmath}
Here, the second line gives the conventional estimator for ${\cal C}$ using the ensemble covariance matrix, assuming the $N$ samples are independent. Throughout this document we will assume the covariance matrix is not known {\it a priori} and must therefore be estimated from the data using this or some other means. As mentioned above, we label the quantity given in  Eq.~\ref{eq-q2-def} as $q^2$ rather than the conventional $\chi^2$ to avoid confusion in cases where the $\chi^2$ distribution is not appropriate. 

Aside from the fit parameters, we are typically also interested in estimating the statistical error, i.e. the error-on-the-mean of our result, and the goodness-of-fit to check our model assumptions. Below we briefly revisit these concepts and the strategies by which they can be computed.

\subsection{Statistical error and bootstrap resampling}\label{sectionOrderNbias}

The error-on-the-mean is defined as the expectation value of the deviation of the ensemble mean from the population mean,
\begin{dmath}
\sigma(t) \equiv \sqrt{{\cal C}(t,t)}\,,
\end{dmath}
for which an estimate can be obtained through Eq.~\ref{eq-covmatdef}. Typically we are interested in complicated combinations of $C(t)$ -- for example, the result of fitting the ensemble means to some model -- for which the propagation of the statistical error becomes rather cumbersome. To simplify the error propagation it is common to apply a {\it resampling} procedure, either bootstrap~\cite{10.1214/aos/1176344552} or jackknife~\cite{10.1093/biomet/43.3-4.353,10.1214/aoms/1177706647}. Both procedures can be described as generating a set of $R$ ``resampled ensembles'' which serve as a proxy for independent experiments, and the error is obtained from the size of the fluctuations in the derived quantity over these resampled ensembles. For a formally-correct application of resampling in the context of a fit, it is necessary to compute the covariance matrix separately for each resampled ensemble as one would do were they truly independent, although in cases where the fluctuations are demonstrably small it can be sufficient to fix the matrix to a single estimate. 

In this document we will focus on (non-parametric) bootstrap resampling, which is more generally applicable than the jackknife, being capable of estimating quantities other than just the statistical error. Bootstrapping generates $R=N_{\rm boot}$ resampled ensembles by randomly drawing $N$ values from the original ensemble according to a uniform distribution (with replacement). For a given property of the population, the bootstrap provides a plug-in estimator by treating each resampled ensemble as if it were a true, independent experiment. The distribution of a measured quantity over the resampled ensembles defines an ``empirical distribution" whose properties mirror those of the true population up to a finite-sample bias whose origin is described below.

For example, for the statistical error we compute the mean on each resampled ensemble of index $b$,
\begin{dmath}
\bbar{C}^{*,b}(t) \equiv \frac{1}{N}\sum_{i=1}^N C_i^{*,b}(t)\,
\end{dmath}
where the $*$ indicates the data belong to a resampled ensemble. From these we compute the variance over the resampled ensembles (i.e. over the empirical distribution),
\begin{dmath}
\sigma^2(t) \approx  \bigtrueavg{ \Big( \bbar{C}^*(t) - \trueavg{\bbar{C}^*(t)}_R \Big)^2 }_R\,,
\end{dmath}
where $\trueavg{\phantom{.}}_R$ indicates the average over the population of bootstrap ensembles, e.g.
\begin{dmath}
\trueavg{\bbar{C}^*(t)}_R \equiv \frac{1}{N_{\rm boot} }\sum_{b=1}^{N_{\rm boot}} \bbar{C}^{*,b}(t)\,.
\end{dmath}
In principle $N_{\rm boot}$ must be large, although typically values of 500-1000 suffice.
 
It may be useful to review the argument upon which the bootstrap is based.  Both the jackknife and bootstrap resampling schemes rely on the large number of samples that make up a single resampled ensemble.  For large $N$ the resampled ensembles are highly similar.  The variations from one resampled ensemble to the next are small and become smaller as the number of samples in each ensemble increases.  Consider a general quantity $\mathcal{Q}$ that is a function of averages of products of the samples $C_i(t)$ over a single ensemble.   For a resampled bootstrap ensemble $b$
\begin{equation}
\mathcal{Q}^{*,b} = \mathcal{Q}\left(\sum_{i_1=1}^N C^{*,b}_{i_1}(t_{1,1}),\sum_{i_2=1}^N C^{*,b}_{i_2}(t_{2,1}) C^b_{i_2}(t_{2,2}),
\sum_{i_3=1}^N C^{*,b}_{i_3}(t_{3,1}) C^{*,b}_{i_3}(t_{3,2}) C^{*,b}_{i_3}(t_{3,3}), \ldots \right)
\end{equation}
where $\{t_{i,j}\}$ represent different choices for the set of times $t$ at which the correlation functions $C(t)$ are measured.  Here the quantity $\mathcal{Q}$ could be a simple average of the correlation function C(t) at a single time or it could be the quantity $q^2$ defined in Eq.~\eqref{eq-q2-def} for fixed parameters $\vec a$ or even one of the $A$ parameters $a_k$ which minimizes $q^2$.

If we wish to determine the probability distribution function for the quantity $\mathcal{Q}$ as the ensemble on which it depends varies over the universe of independent experiments, we might consider the $n^{th}$ moment:
\begin{equation}
\Delta^n \mathcal{Q} = \bigtrueavg{\Bigl(\mathcal{Q} - \trueavg{\mathcal{Q}}\Bigr)^n}
\label{eq:PDF-moment}
\end{equation}
and the extent to which this quantity is accurately reproduced by the average over $N_{\mathrm{boot}}$ bootstrapped ensembles created from a single ensemble.  Since the arguments of $\mathcal{Q}$ are ensemble averages, their fluctuations as that ensemble is varied are of order $1/\sqrt{N}$ suggesting that Eq.~\eqref{eq:PDF-moment} might be evaluated by performing a Taylor series expansion in the fluctuations of these averaged quantities about their true average.  In the case of the bootstrap estimate of this $n^{th}$ moment we would expand in the fluctuations of the bootstrap ensemble averages from a single ensemble about the average for that ensemble and then average over the universe of such ensembles.   

The difference between the average of independent ensembles and average of bootstrap-generate ensembles can be recognized by considering the product of two averages that will occur in such a Taylor expansion of the quantity $\mathcal{Q}$ in fluctuations about the true or an ensemble average:
\begin{equation}
\bigtrueavg{\left(\sum_{i=1}^N C_i(t)\right) \left(\sum_{j=1}^N C_j(t')\right)}
 = \frac{1}{N_{\mathrm{boot}}}\sum_{b=1}^{N_{\mathrm{boot}}}
  \bigtrueavg{\left(\sum_{i=1}^N C^{*,b}_i(t)\right) \left(\sum_{j=1}^N C^{b,j}_j(t')\right)}
    + O\left(\frac{1}{N}\right).
\label{eq:1-over-N}
\end{equation}
As indicated by this equation, the difference between these two averages will be of order $1/N$.  In the left hand side of Eq.~\eqref{eq:1-over-N} for each $C_i(t)$ in the left-hand factor, because the $N$ samples $C_j(t)$ are independent, there will be only a single term in the right-hand factor with which it is correlated: that term with $j=i$.  However, on the right-hand side of that equation, for each $C^b_i(t)$ in the left-hand factor there may be more than one term in the right-hand factor which is identical to $C^b_i(t)$ because the samples of a bootstrap ensemble may not be all independent -- some may be repeated.  For a given pair $C^{*,b}_i(t) C^{*,b}_j(t')$ on the right-hand side for $i\ne j$ there is a $1/N$ probability that the $i^{th}$ and $j^{th}$ samples are the same, $C^{*,b}_i = C^{*,b}_j$ while on the left-hand side this probability is zero.  This is the origin of the $1/N$ discrepancy between the true average on the right-hand side and the true average of the bootstrap estimate on the left.

Of course the larger the number of factors of possibly identical samples which appear in higher powers in this Taylor series expansion the more of these $1/N$ errors will occur.  While the error in the bootstrap estimate will still fall as $1/N$ the prefactor of $1/N$ will grow, suggesting that this argument requires the Taylor series expansion upon which it is based converge quickly.  Thus, we expect that the bootstrap method can be used to determine the probability distribution function for a quantity defined through an ensemble average with a bias falling as $1/N$.

\subsection{Goodness-of-fit}
\label{sec-goodness-of-fit}

The goodness-of-fit is defined through the {\it null distribution} ${\cal P}_{\rm null}$, which describes the distribution of the minimum of $q^2$, $q^2_{\rm min}$, over many independent experiments, under the assumption that the model is a correct description of the data and thus has no systematic deviation. We emphasize that the definition of ``experiment'' is context-dependent, describing the specific procedure that the experimenter is employing; in our lattice calculation it should be considered as repeating the same analysis procedure on each of many independently-generated ensembles, {\it including} independently estimating the covariance matrix upon each ensemble. As such, both the fluctuations of the ensemble means about the population mean and those of the ensemble covariance matrix about the population matrix impact the form of the null distribution. 

To describe the goodness-of-fit we typically compute the (one-sided) p-value,
\begin{dmath}
p(q^2_{\rm min}) = \int_{q^2_{\rm min}}^{\infty} dq^{\prime\ 2}{\cal P}_{\rm null}(q^{\prime\ 2}) \label{eq-pvaluedef}
\end{dmath}
which represents the probability of obtaining a {\it worse} fit than the one performed. Assuming the null distribution is a valid description of the fluctuations of the data around the model function, a low value represents either an ``unlucky draw'' from the population distribution or that the model itself is not a good description of the data.

Note, below and in the remainder of this document, we will typically be discussing only the minimized value of $q^2$, hence we will drop the subscript ``min" unless there is an ambiguity.

In practice we typically cannot repeat our experiments many times, and so we must determine the appropriate null distribution through some other means. Assuming normally-distributed data, the null distribution can be determined analytically, giving the Hotelling $T^2$ distribution~\cite{hotelling1931} with probability density function (PDF):
\begin{dmath}
{\cal P}_{T^2}(q^2; K,n) = \rho  {\cal P}_{F}( \rho q^2\ ;\  K,\  n-K+1 )\,,
\end{dmath}
where $n=N-1$ for $N$ samples, $K=T-A$ degrees of freedom, $T=t_{\rm max}-t_{\rm min}+1$,
\begin{dmath}
\rho = \frac{n-K+1}{nK}\,,
\end{dmath}
and ${\cal P}_{F}$ is the PDF of Snedecor's $F$ distribution,
\begin{dmath}
{\cal P}_{F}(x; d_1, d_2) =  \frac{  \sqrt{  \frac{ (d_1 x)^{d_1} d_2^{d_2}  }{ (d_1 x + d_2)^{d_1+d_2} } }    }{ xB\left(\frac{d_1}{2}, \frac{d_2}{2} \right)  }\,,
\end{dmath}
where $B$ is the beta function. 

It is not difficult to see that the Hotelling $T^2$ distribution converges to the $\chi^2$ distribution for $K$ degrees of freedom in the limit of large $N$. To do so, we can consider the behavior of $q^2$ given in Eq.~\eqref{eq-q2-def} in this limit. When considering the large-$N$ behavior of this quantity it is convenient to first multiply both the numerator and denominator by a factor of $N$. Upon doing so, the two factors that make up the numerator take on the form $\sqrt{N}\left( \bar{C}(t) - f(t; \vec a) \right)$. If the data are null consistent, the fit function exactly describes the population mean in the large-$N$ limit: $f(t;\vec a) \xrightarrow{N\to\infty} \llangle C(t) \rrangle$. The resulting difference, $\bar{C}(t) - \llangle C(t) \rrangle$,  quantifies the fluctuation of the ensemble mean about the population mean, the size of which falls like $1/\sqrt{N}$ due to the central limit theorem. As such, $\sqrt{N}\left( \bar{C}(t) - f(t; \vec a) \right) \sim {\cal N}(0,\Sigma(t))$ at large $N$, where $\Sigma(t)$ is an intrinsic property of the underlying distribution, hence it remains finite in this limit. Since the interpretation and assumed behavior of the numerator is the same whether we are considering the $\chi^2$ or Hotelling distributions, it is not the source of the difference between these two descriptions of $q^2$.

Likewise, when multiplied by $N$, the denominator in Eq.~\eqref{eq-q2-def} also remains finite in the limit of large $N$.  As defined in the first line of Eq.~\eqref{eq-covmatdef}, ${\cal C}(t,t')$ provides the size of expected statistical fluctuation of the numerator from one ensemble to another.  It is defined precisely as a population average.  However, the second line in Eq.~\eqref{eq-covmatdef} replaces ${\cal C}(t,t')$ by its estimate from the ensemble being studied. Multiplying the estimate by $N$ gives the ensemble (co-)variance, which quantifies the fluctuations of individual samples about the ensemble mean. At large $N$ the ensemble mean approaches the population mean, hence the limiting behavior of the denominator multiplied by $N$ describes the second moments of the population, an intrinsic property of the underlying distribution without any dependence on $N$. Therefore, the estimate of the population property in the second line in Eq.~\eqref{eq-covmatdef} becomes exact in this limit. However, at large but finite $N$, this estimate has statistical errors which fall as $1/\sqrt{N}$ and it is the effect of these fluctuations on the $q^2$ distribution which the Hotelling $T^2$ distribution captures. Given that it is only the consideration of these fluctuations in the denominator, which fall like $1/\sqrt{N}$, that distinguishes the Hotelling from the $\chi^2$ distribution, it is clear that in the limit of large $N$ these two distributions must be the same.

The conclusion that the $\chi^2$ distribution is the large $N$ limit of the null distribution does not rely on any assumptions on the distribution of the underlying data, since the fluctuations of the ensemble means are always normal by the central limit theorem. However, for a lattice calculation with $N{\sim}{\cal O}(100-1000)$ samples, one must be careful with assuming the $\chi^2$ distribution as the corrections associated with the fluctuations of the covariance matrix can become large rather quickly when the number of degrees of freedom $K$ is large. This can be seen, for example, in the convergence of the mean of the $T^2$ distribution to that of $\chi^2$, which goes like $K^2/N$. As such, for a typical lattice simulation we can have significant deviations from the $\chi^2$ distribution -- and hence invalid p-values -- even for modest numbers of degrees of freedom. 

Unfortunately, it is common for a lattice simulation to have insufficient $N$ to obtain a reliable determination of the covariance matrix, resulting in unexpectedly large statistical errors on the fit result. One cause of this breakdown is that the covariance matrix may be large enough that its fluctuations over the resampled ensembles -- or indeed over independent experiments -- dominate the error on this result. We discuss an example of this in greater detail in Section~\ref{sec-autocorr-ignoring-covmat}. This enhancement of the statistical error can typically be remedied to a degree by restricting the fit range, at the cost of excluding valuable data. Another cause is the dominance in $q^2$ of the contributions associated with poorly determined small eigenvalues, an example of which is described in Ref.~\cite{Yoon:2011wdw}. For this, and also for the previous case, a common workaround is to manipulate the covariance matrix in some way, for example by truncating the eigenspectrum or even zeroing all off-diagonal elements (an uncorrelated fit). In doing so, one regains control over the covariance matrix at the cost of potentially larger statistical errors than one would obtain with a true (stable) correlated fit. Perhaps more importantly, one also loses the ability to estimate the goodness-of-fit, as modifying the analysis procedure by manipulating the covariance matrix changes the definition of $q^2$ such that the analytical results are no longer applicable. The procedure described in this document remedies the latter issue.

Another common issue with lattice calculations is the presence of autocorrelations in the Markov chain Monte Carlo data, which are not accounted for in the analytical formulae for the null distributions, and so inaccurate p-values will be obtained if applied directly. This issue is commonly addressed by binning, i.e. pre-averaging over consecutive blocks of data within the ensemble prior to performing the fit analysis. With sufficient bin sizes, the binned samples are effectively independent, and moreover are normally distributed by the central limit theorem, hence in that limit the Hotelling $T^2$ distribution once again becomes the correct null distribution. Unfortunately, as binning reduces the number of samples available to compute the covariance matrix, this approach has the effect of amplifying further the fluctuations in the covariance matrix to the point where the procedure can break down and modifying the definition of $q^2$ once again becomes necessary. In Section~\ref{sec-autocorrelated-data} we will demonstrate the efficacy of our new procedure in the context of autocorrelated data.

\subsection{Toy model demonstrations and obtaining the ``true" null distribution}
\label{sec-toy-model-demos}

In order to explore approaches beyond the conventional analytical description of the null distribution, it is necessary to define a procedure for determining the ``true" null distribution and a toy model framework in which we can demonstrate and test our approach.

To approximate the null distribution, we will generate many statistically independent ensembles of toy data. Each ensemble consists of $T$ ``timeslices," for each of which $N$ measurements are drawn, described by the model function plus some random noise. For each ensemble, we then fit model parameters $\vec a$ by minimizing $q^2$. The resulting list of $q^2$ is thus sampled directly from the null distribution and represents its empirical probability distribution. We experiment with values of $N$ in the range of 100-1000 to match the typical amount of data available for a lattice calculation. This is also within the range of $N$ for which the fluctuations of the covariance matrix across ensembles impact the shape of the null distribution.

Although many different model functions $f(t; \vec a)$ are possible, all of the essential features of the process are revealed by fitting to a constant; that is, $f(t; a) = a$. Thus, unless otherwise stated, we will use this model function for our demonstrations throughout this document. The value of this constant is irrelevant to the end result, hence we will take it to be $0$ for the purposes of data generation. Thus, the ensembles consist entirely of random noise with arbitrary distributions (of mean zero), correlations and potentially autocorrelations. 

\begin{figure}[t]
\centering
\includegraphics[scale=0.7]{"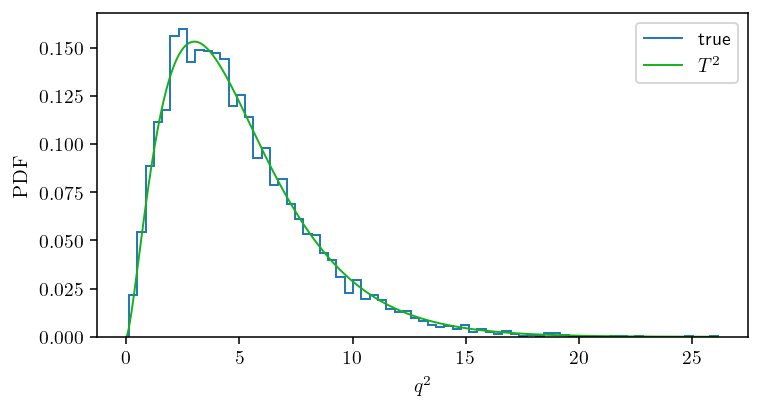"}
\caption{A histogram of the $q^2$ values obtained from correlated fits to toy data generated independently for each timeslice from a normal distribution, $C_i(t)\sim {\cal N}(0,1)$ with $N=800$ and $T=6$ ($K=5$) (blue) compared to the expected probability density of the null distribution (green) obtained from $T^2$.}
\label{fig:null_gaussian}
\end{figure}

\begin{figure}[t]
\centering
\includegraphics[scale=0.7]{"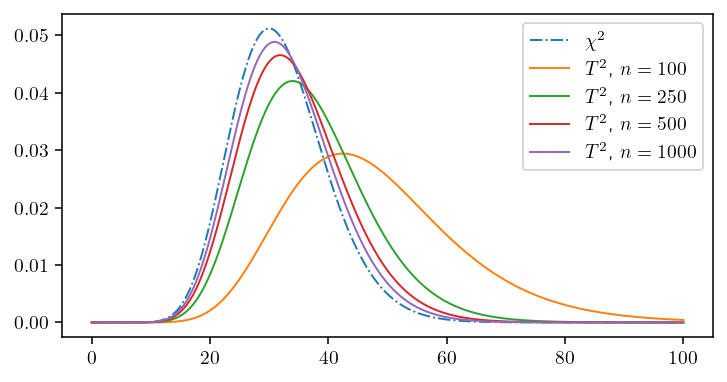"}
\caption{Histograms of the $\chi^2$ and $T^2$ distributions for $K=32$ degrees of freedom and different ensemble sizes, $N$,  demonstrating the convergence of the two at large-$N$.}
\label{fig:chi2vst2}
\end{figure}

For normally-distributed data, the null distribution for a correlated fit is described analytically by the $T^2$ distribution. We demonstrate this in Fig.~\ref{fig:null_gaussian} using our toy model for an ensemble of size $N=800$ and $T=6$ timeslices ($K=5$). For reference, Fig.~\ref{fig:chi2vst2} shows how the $T^2$ distribution differs from the more familiar $\chi^2$ distribution for $K=32$ degrees of freedom (to emphasize the difference) and different values of $N$. We can see that the distributions converge as $N$ grows.

\begin{figure}[t]
\centering
\includegraphics[scale=0.7]{"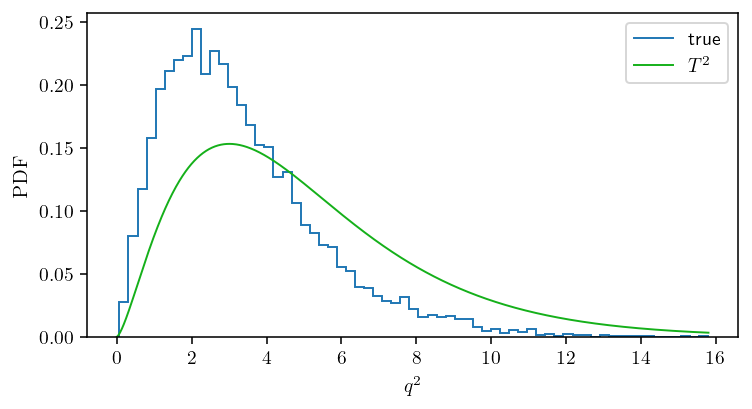"}
\caption{A histogram of the $q^2$ values obtained from uncorrelated fits to the same toy data used in Fig.~\ref{fig:null_gaussian} 
compared to the probability density of the $T^2$ distribution, demonstrating the inapplicability of the $T^2$ null distribution for obtaining the goodness-of-fit for uncorrelated fits. }
\label{fig:null_uncorrelated}
\end{figure}

We also want to estimate the null distribution in situations where we manipulate the covariance matrix in some way before fitting model parameters. Fig.~\ref{fig:null_uncorrelated} shows the null distribution in the case of an uncorrelated fit to the same data; as expected, the null distribution differs significantly from $T^2$.

\section{Bootstrap estimation of the null distribution}
\label{sec-bootstrap-null-dist}

\subsection{Description of the bootstrap approach}
\label{sec-bootstrap-approach-descr}

We now introduce a procedure to estimate the null distribution directly from our data via bootstrap resampling. For a bootstrap resampled ensemble $b\in \{1\ldots N_{\rm boot}\}$ we define,
\begin{equation}
\bbar{C}^{*,b}(t) = \bbar{C}(t) + \bar\epsilon^{*,b}(t)\label{eq-bootstrpdev}
\end{equation}
where as before $\bbar{C}^{*,b}(t)$ is the mean of the resampled ensemble and $\bbar{C}(t)$ that of the original ensemble. The bootstrap analogue of the central limit theorem implies that over the bootstrap ensembles, 
\begin{equation}
\bar\epsilon^{*\,b}(t){\sim}{\cal N}(0,\sigma^2(t))\label{eq-bootstrpcnlm}
\end{equation}
up to a bias typically ${\cal O}(1/N)$. To a similar level of accuracy, the distribution of covariance matrices over bootstrap ensembles also reflects that of the population as previously discussed.

We also define the fluctuations of the original ensemble means about the fitted model,
\begin{equation}
\bbar{C}(t) = f(t; \vec a) + \bar\xi(t)\,,
\end{equation}
where $\vec a$ are the best-fit parameters obtained from the original ensemble. Note that we do not assume any specific distribution for $\bar\xi$. If the model is a perfect description of the population of means (i.e. the null hypothesis holds), then $\bar\xi$ will be symmetrically distributed about zero; however, this is not necessary for the validity of the bootstrap procedure.

Incorporating the above gives
\begin{equation}
\bbar{C}^{*,b}(t) = f(t; \vec a) + \bar\xi(t) + \bar\epsilon^{*\,b}(t) \label{eq-bootstrpdev2}
\end{equation}
which are distributed over resampled ensembles as ${\cal N}(f(t; \vec a)+\bar\xi(t),\sigma^2(t))$.

The definition of the null distribution is the distribution of $q^2$ for data that are distributed about the model. The bootstrap means in Eq.~\ref{eq-bootstrpdev2} are not suitable for this application due to the extra term, $\bar\xi(t)$, which remains fixed over all resampled ensembles. Fortunately, we can easily correct for this and thereby {\it impose} the null hypothesis on the bootstrap ensembles by applying a {\it recentering} to each sample:
\begin{dmath}
\widetilde C^{*\,b}_i(t) = C^{*,b}_i(t) - \bar\xi(t) = C_i^{*,b}(t) - \bar C(t) + f(t; \vec a) \label{eq-recentering}
\end{dmath}
such that, from Eq.~\ref{eq-bootstrpdev},
\begin{dmath}
\bbar{\widetilde C}^{*\,b}(t) = f(t; \vec a) + \bar\epsilon^{*\,b}(t)
\label{eq-recentered-ens-mean}
\end{dmath}
which are distributed as ${\cal N}(f(t; \vec a),\sigma^2(t))$. Note that as the shift above is applied uniformly to all samples in each bootstrap ensemble, the bootstrap distribution of covariance matrices -- which reflects that of the population -- is unaffected. 

Now that we can generate data that are appropriately distributed about the model, we need only apply our measurement procedure -- i.e. perform a fit to the recentered data -- for a large number, $N_{\rm boot}$, of bootstrap ensembles and construct an empirical distribution (histogram) of $q^2$ values, approximating ${\cal P}_{\rm null}(q^{\prime\ 2})$, from which the p-value can be estimated via Eq.~\ref{eq-pvaluedef}. Alternatively, one can simply count the number of bootstrap values above $q^2_{\rm min}$ and divide by $N_{\rm boot}$ to obtain the plug-in estimate of the p-value. Note that in practice the resulting estimate, particularly for low-$q^2$ where the sensitivity to the tails of the distribution is greatest, will contain an  error due to being estimated at finite $N_{\rm boot}$. This can of course be reduced simply by increasing $N_{\rm boot}$ at the cost of increasing analysis time, or else one might imagine a more sophisticated smooth interpolation scheme; however, this source of error is typically much smaller than the bias or the ``statistical error" discussed below, and can therefore be neglected.

\subsection{Toy model demonstration}
\label{sec-toy-model-demo-null-consistent}

\begin{figure}[h]
\centering
\includegraphics[scale=0.7]{"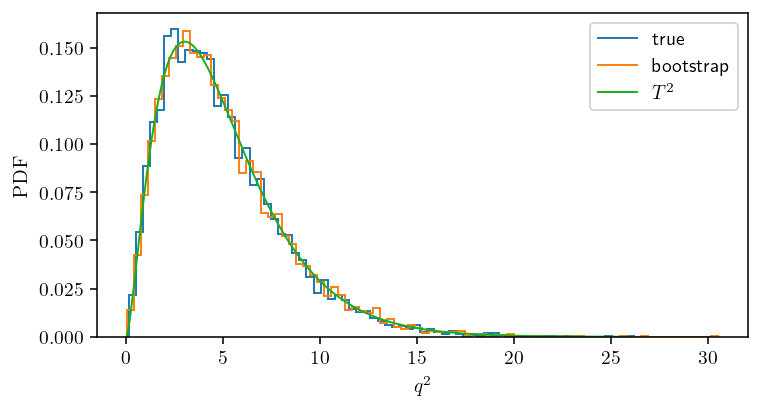"}
\caption{A histogram of the bootstrap null distribution for $N=100$, $L_T=6$. Each configuration is generated from a multivariate normal distribution with mean $\mu(t)=0$ and input covariance matrix $\Sigma(t_1,t_2)=1/2 + \delta_{t_1,t_2}$.}
\label{fig:bootstrap_gaussian}

\end{figure}

\begin{figure}[h]
\centering
\includegraphics[scale=0.7]{"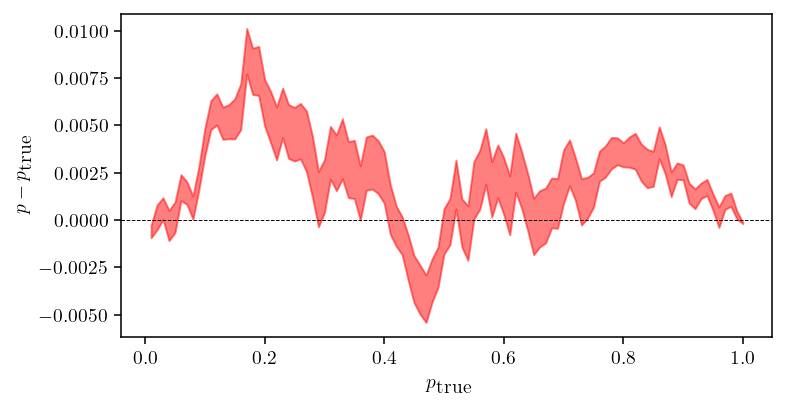"}
\caption{Difference in the p-value obtained from the bootstrap procedure vs. the exact one computed from $T^2$ and from multivariate normal data as in Fig.~\ref{fig:bootstrap_gaussian}.}
\label{fig:pdiff_gaussian}
\end{figure}

We now demonstrate the bootstrap procedure on our toy models. When the data consist of Gaussian noise, we know that the null distribution is described by $T^2$. In Fig.~\ref{fig:bootstrap_gaussian} we demonstrate this for toy data with a non-trivial covariance structure by employing the multi-variate normal distribution, again fitting to a constant, where we observe the bootstrap procedure replicates both the ``true" distribution and the analytic result. In Fig.~\ref{fig:pdiff_gaussian}, we plot the resulting discrepancy between the p-value we obtain by bootstrapping and the p-value that we would have obtained analytically from $T^2$. The interpretation and means of estimating the error bars in this plot are described in the next section. We note that in this case the worst error on the p-value is around $0.3\%$, indicating good agreement in this situation where we have the exact form of the null distribution. 

\begin{figure}[t]
\centering
\includegraphics[scale=0.7]{"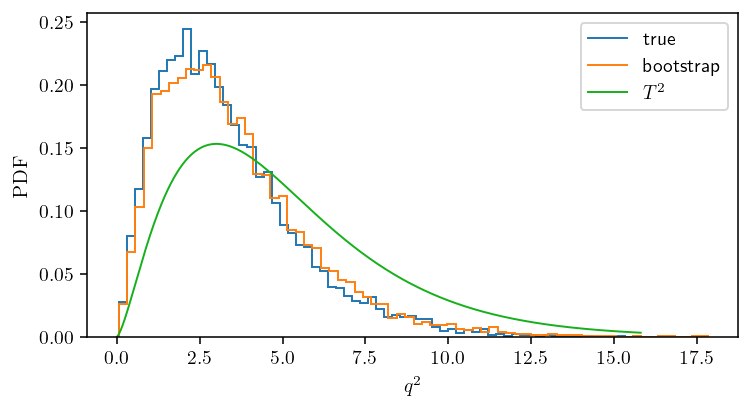"}
\caption{Bootstrap distribution of $q^2$, for an uncorrelated fit to multivariat normal data as in Fig.~\ref{fig:bootstrap_gaussian}.}
\label{fig:bootstrap_uncorrelated}

\end{figure}

\begin{figure}[t]
    \includegraphics[width=0.48\textwidth]{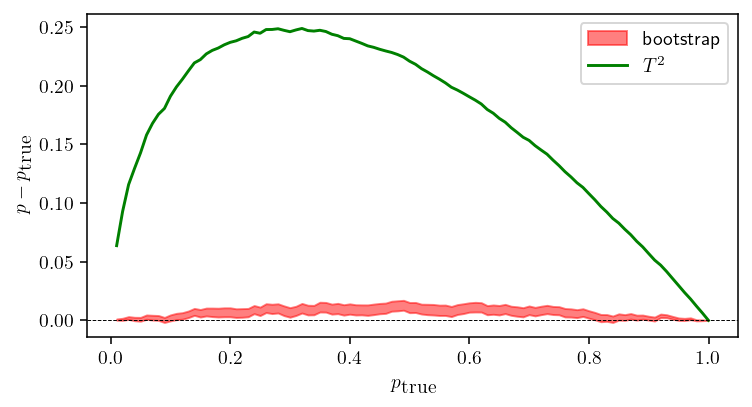}
    \includegraphics[width=0.48\textwidth]{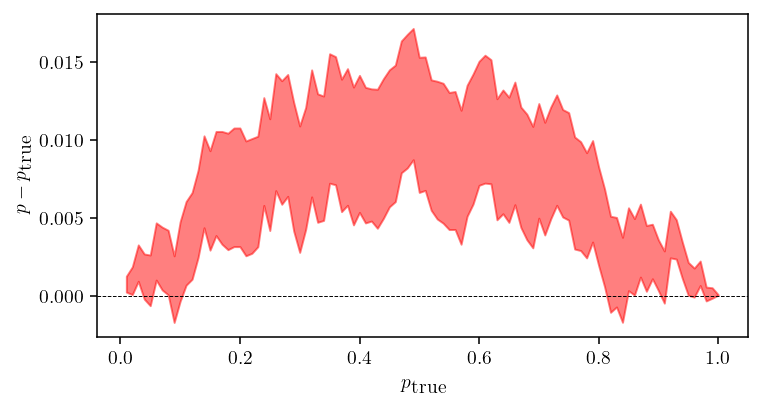}
  \caption{The difference between the p-value obtained via the bootstrap procedure and the true value estimated from many independent ensembles, for uncorrelated fits to data generated as per Fig.~\ref{fig:bootstrap_gaussian}. The plot on the left shows the deviation including the corresponding $T^2$ p-value, and the plot on the right shows just the bootstrap result. \label{fig:pdiff_uncorrelated} }
\end{figure}

If we perform an uncorrelated fit to the same data, the null distribution deviates significantly from $T^2$. Nevertheless, from a single ensemble, the bootstrapping procedure allows us to estimate a null distribution which agrees well with the true distribution obtained using many independent independent ensembles. In Fig.~\ref{fig:bootstrap_uncorrelated} we show the distribution of $q^2$ for an uncorrelated fit, with the $T^2$ distribution for comparison. In Fig.~\ref{fig:pdiff_uncorrelated}, we plot the difference between the bootstrapped p-value (with error bars) and the true p-value, as well as the difference between the p-value obtained from $T^2$ and the true p-value. Again we observe that the bootstrap procedure reliably estimates the null distribution even in a case where no analytic description is known.

\subsection{Statistical error on the p-value}
\label{sec-p-value-stat-err}

The p-value we obtain using the bootstrap procedure for any given value of $q^2$ is dependent on the original ensemble that we resample from. This reflects the fact that the null distribution in a general case depends upon the parameters of the function describing the population means, and with a single ensemble we have no means of estimating those parameters other than from the best fit to our original ensemble. Assuming the ensemble is null consistent, the best-fit parameters will deviate from those of the population due to statistical effects. A second source of deviation is due to the variation in the degree to which the fluctuations of the samples in the original ensemble are reflective of those of the population, as these affect both the fluctuations under the bootstrap of the covariance matrix in the denominator of $q^2$ and of the means in its numerator, as well as the values of the empirical population covariance matrix itself. Both of these effects are statistical in nature and vanish at large $N$, hence the resulting fluctuation in the p-value for a given $q^2$ across null distributions estimated using different original ensembles can be thought of as a kind of statistical error or standard error (the fluctuation of an estimate about the mean of the distribution from which it was sampled).

In this document we often wish to compare, as in Fig.~\ref{fig:pdiff_gaussian}, the bootstrap estimate of the p-value to the true p-value. The purpose of these plots is to illustrate the {\it bias} of the bootstrap estimate, i.e. the deviation of the typical estimate from its true value, as a function of $q^2$. We estimate this typical value by repeating the bootstrap procedure over $M$ different, independent original ensembles and averaging the result:
\begin{dmath}
\trueavg{p_B}\approx \frac{1}{M}\sum_{I=1}^M p_B^I
\end{dmath}
where $p_B^I$ is the bootstrap p-value estimate obtained from original ensemble $I$. As performing the bootstrap procedure has some non-trivial numerical cost, we typically use $M{\sim}{\cal O}(20)$. An uncertainty in our value for this bootstrap mean estimate arises from using this relatively low value of $M$, hence we also include on these plots an error bar reflecting the error-on-the-mean obtained in the usual way:
\begin{dmath}
\sigma^2_{\trueavg{p_B}} \approx {\rm var}(p_B^I)/(M-1)\,.
\end{dmath}

The error bar on the plots of the bias is distinct from the ``statistical error" described above.  The statistical error reflects the size of the fluctuations in the bootstrap estimate as we vary the original ensemble,
\begin{dmath}{
\sigma^2_{\rm stat} = \bigtrueavg{(p_B - \trueavg{p_B})^2} \approx {\rm var}(p_B^I)
}
\label{eq:stat_error}
\end{dmath}
again estimated over $M$ different original ensembles.  Note the absence of the factor of $1/(M-1)$ in Eq.~\eqref{eq:stat_error}.  When quoting a p-value obtained from the bootstrap procedure it is this error that should be associated with that quoted p-value. 

\begin{figure}[t]
\centering
\includegraphics[width=0.48\textwidth]{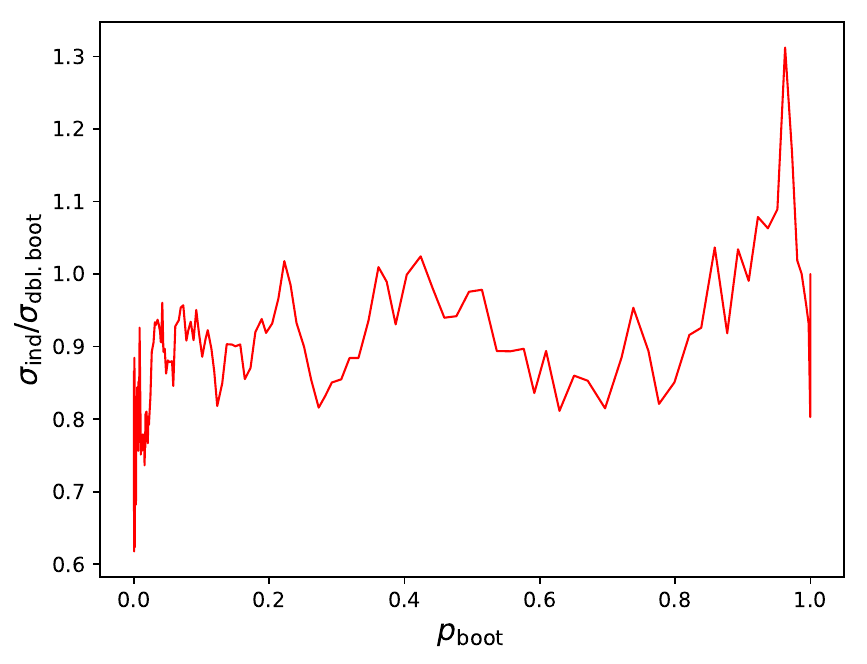}
\includegraphics[width=0.48\textwidth]{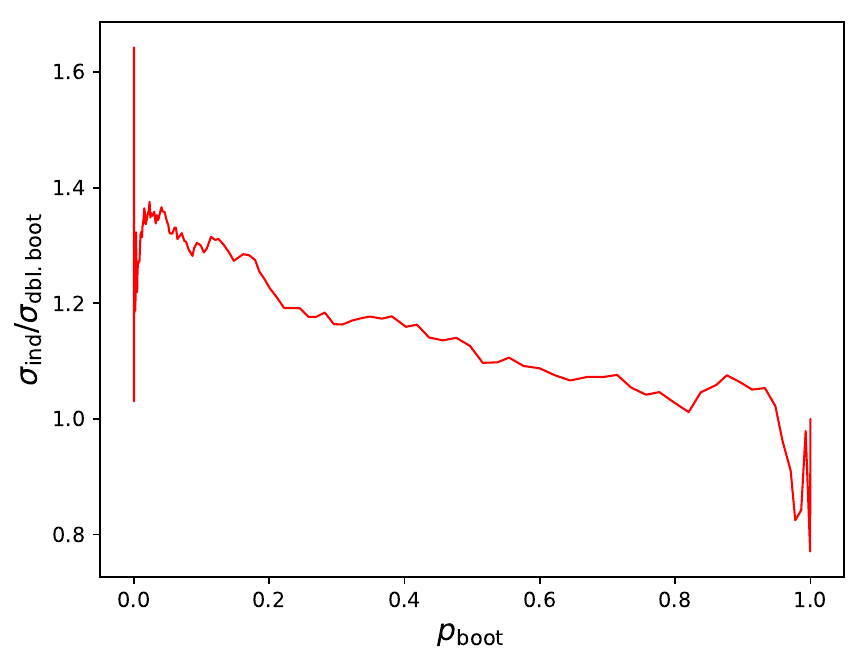}
\caption{The ratio of the ``statistical error" on the bootstrap p-value $p_{\rm boot}$ as estimated via $M=50$ independent original ensembles to that estimated via $M=50$ bootstrap resampled ensembles derived from a single original. The left and right figures show this ratio vs. $p_{\rm boot}$ for a correlated and an uncorrelated fit, respectively, to data generated as described in Fig.~\ref{fig:bootstrap_gaussian}. The uneven sampling frequency is due to sampling uniformly over $q^2$ and not $p_{\rm boot}$. \label{fig-stat-err-demo} }
\end{figure}

In practice we do not typically have $M$ different original ensembles with which to estimate the statistical error. Our proposal is to replace these with $M$ bootstrap resampled ensembles derived from a single original ensemble. In summary, we propose the following steps to provide an estimate of the statistical error on the bootstrap p-value obtained through the method described in Sec.~\ref{sec-bootstrap-approach-descr}:
\begin{enumerate}

\item Generate $M$ ensembles from the original using bootstrap resampling.
\item Use the algorithm described in Sec.~\ref{sec-bootstrap-approach-descr} to obtain separate estimates of the null distribution from each of these ensembles.
\item For each of these estimated null distributions, compute the p-value for the $q^2$ obtained from the fit to the original ensemble.
\item Use the variance of the resulting $M$ p-values as an estimate for the square of the statistical error on the original p-value.
\end{enumerate}
Of course this ``double-bootstrap" error estimate will itself contain the usual bootstrap bias and may not fully encapsulate the variation in the degree of fluctuation. In Fig.~\ref{fig-stat-err-demo} we compare the double-bootstrap error estimate to that obtained from $M=50$ truly independent ensembles. For the correlated fit we observe good agreement, however for the uncorrelated fit we found the double-bootstrap underestimated the error by 20-30\% in the lower p-value regime. This is likely exacerbated in the chosen example by the strong correlations in the underlying, multivariate normal data. Nevertheless, the estimate serves as a reasonable enough approximation for our purposes (and certainly the best we can achieve given the constraints on obtaining the data).

As to the ``systematic error" on our bootstrap p-value resulting from the finite-$N$ bias discussed in Sec.~\ref{sectionOrderNbias}, it remains a topic of future investigation whether a suitable bootstrap estimation strategy exists. We discuss this bias further in Sec.~\ref{sec-bootstrap-autocorr-ignore-pvalue}.

\subsection{Goodness-of-fit estimation for eigenvalue truncation methods}
\label{sec-eval-trunc-demo}

In some cases, fit instability can be attributed to the presence of a few poorly-resolved low modes of the covariance matrix. Ref.~\cite{Yoon:2011wdw} examines two possible strategies for regaining stability that involve truncating or otherwise manipulating the eigenspectrum of the covariance matrix: the {\it cutoff} method simply removes the offending eigenvalues from the inverse covariance matrix entering the calculation of $q^2$:
\begin{dmath}
\calC_{\rm cut}^{-1} = \sum_{\alpha, \lambda_\alpha > \epsilon} v_\alpha \frac{1}{\lambda_\alpha} v^T_\alpha
\end{dmath}
where $\calC_{\rm cut}^{-1}$ is the resulting truncated inverse-covariance matrix, $v_\alpha$ are the eigenvectors of the original covariance matrix and $\lambda_\alpha$ its eigenvalues, $\epsilon$ is the eigenvalue cutoff, and $\alpha\in\{1..T\}$.  (Note, because the truncation of the eigenspace makes the matrix rank deficient, the corresponding truncated covariance matrix is undefined.)

The second, {\it modified covariance matrix} (MCM), strategy investigated in Ref.~\cite{Yoon:2011wdw} proceeds via the eigenspectrum of the correlation matrix defined as
\begin{dmath}
\tilde \calC_{ij} = \frac{\calC_{ij}}{\sigma_i \sigma_j}
\end{dmath}
where $\sigma_i =\sqrt{\calC_{ii}}$ with $\calC_{ij}$ the original covariance matrix. The correlation matrix has $\tilde \calC_{ii}=1$ by definition. One then defines a truncated correlation matrix
\begin{dmath}
\tilde \calC_{\rm trunc} = \sum_{\alpha, \tilde\lambda_\alpha > \tilde\epsilon} \tilde v_\alpha \tilde\lambda_\alpha \tilde v^T_\alpha
\end{dmath}
where $\tilde v_\alpha$ and $\tilde\lambda_\alpha$ are the eigenvectors and eigenvalues of $\tilde \calC$ and $\tilde\epsilon$ the eigenvalue cutoff on the correlation matrix. As before, this matrix is rank deficient; however, this can be remedied by restoring the diagonal elements to unity:
\begin{dmath}
(\tilde \calC_{\rm MCM})_{ij} = \left\{\begin{array}{l|l} 1 & i=j \\ (\tilde \calC_{\rm trunc})_{ij} & i\neq j    \end{array}       \right.\,.
\end{dmath}
One can then obtain a covariance matrix as
\begin{dmath}
(\calC_{\rm MCM})_{ij} = (\tilde \calC_{\rm MCM})_{ij}\sigma_i\sigma_j\,.
\end{dmath}

Both strategies were found to improve the quality of the fit for the specific case analyzed in Ref.~\cite{Yoon:2011wdw}, although it was noted by the authors that these strategies will fail if the eigenmodes being removed are not sufficiently well separated from the others that fluctuations between bootstrap/jackknife resampled ensembles may change the number and character of the eigenmodes removed in each case. However, the major drawback described by the authors was the loss of the ability to assign a goodness-of-fit metric. 

\begin{figure}[t]
\centering
\includegraphics[width=0.49\textwidth]{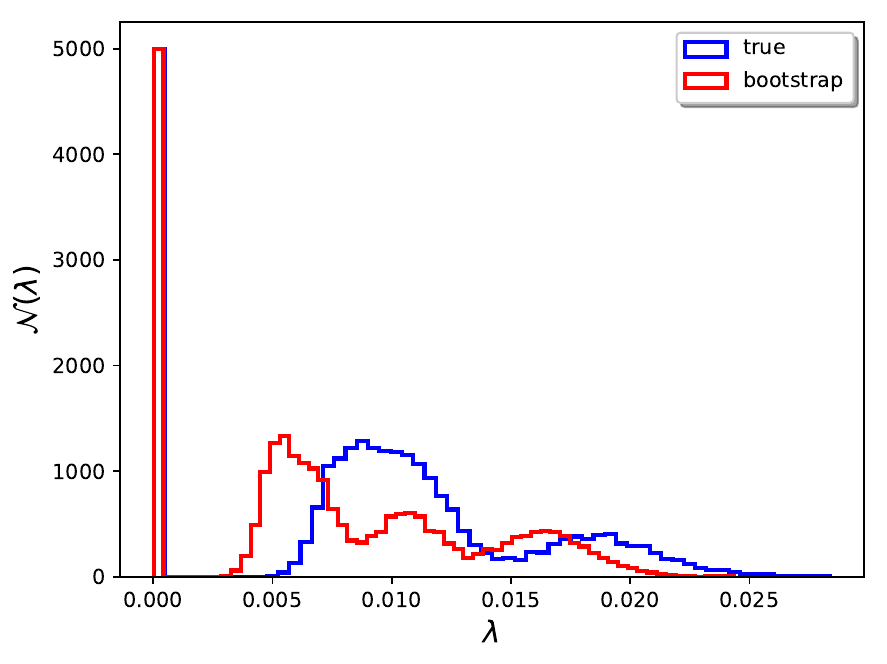}\includegraphics[width=0.49\textwidth]{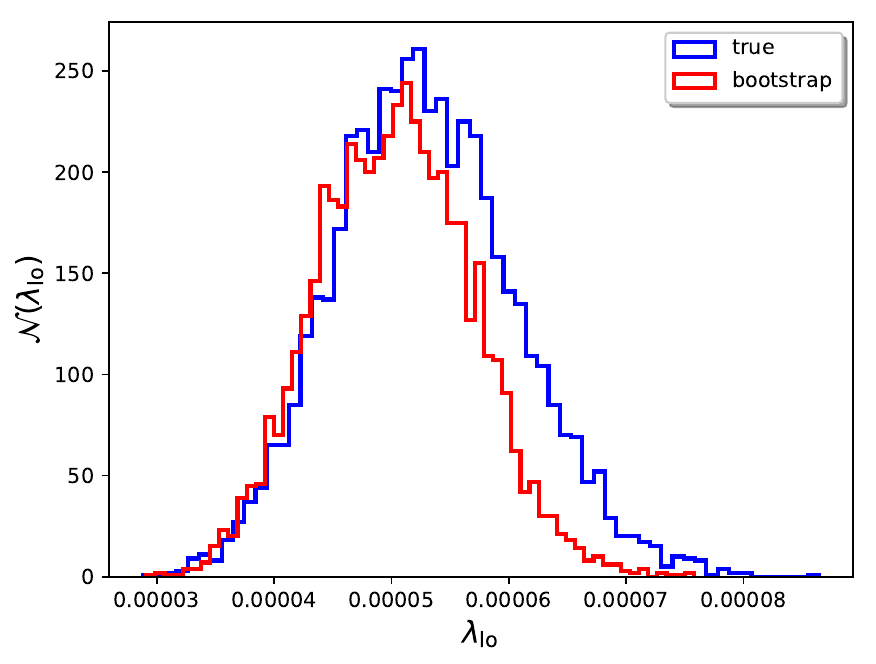}
\caption{Histograms of the complete eigenspectrum (left) and of the lowest eigenvalue (right) for the estimated covariance matrix from ensembles generated as described in Eq.~\ref{eq-mixleft-datagen} with $N=100$, $T=5$, $\alpha_{i<5}=0$, $\alpha_5=0.9$, $\mu=0$ and $\sigma=1$. Results are shown for 5000 indepependently-generated ensembles (``true") and as many bootstrap ensembles (``bootstrap") generated from a fixed original ensemble. \label{fig-covmat-evals-mixleft}  }
\end{figure}

\begin{figure}[t]
\centering
\includegraphics[width=0.49\textwidth]{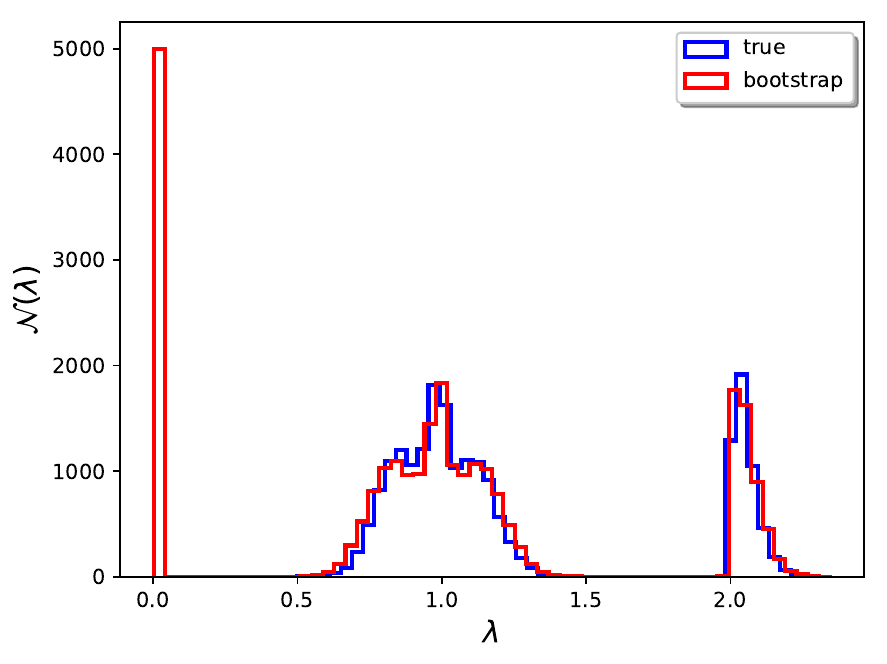}\includegraphics[width=0.49\textwidth]{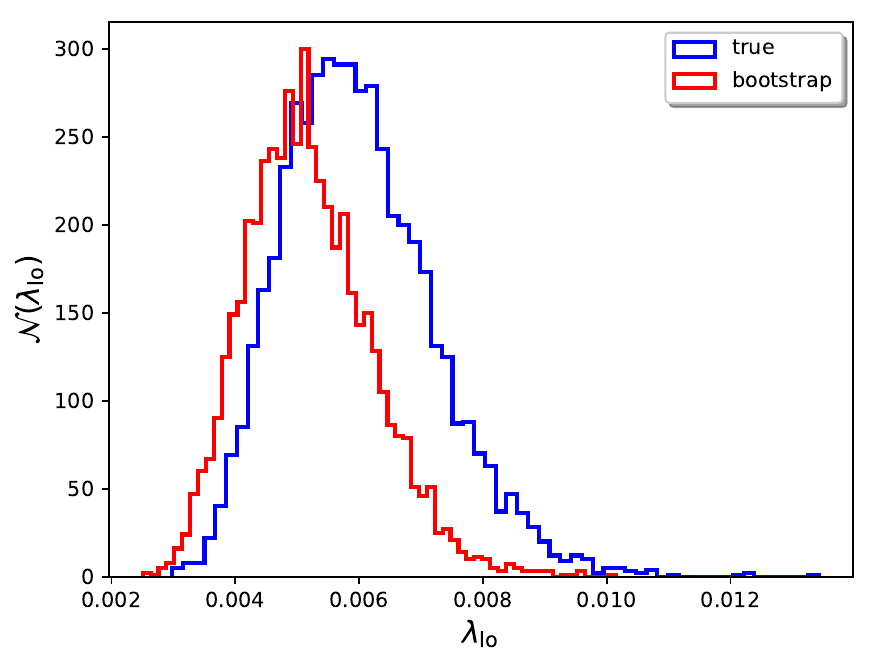}
\caption{Histograms of the complete eigenspectrum (left) and of the lowest eigenvalue (right) for the estimated correlation matrix for the ensembles described in the caption of Fig.~\ref{fig-covmat-evals-mixleft}. \label{fig-corrmat-evals-mixleft}  }
\end{figure}

\begin{figure}[t]
\centering
\includegraphics[width=0.49\textwidth]{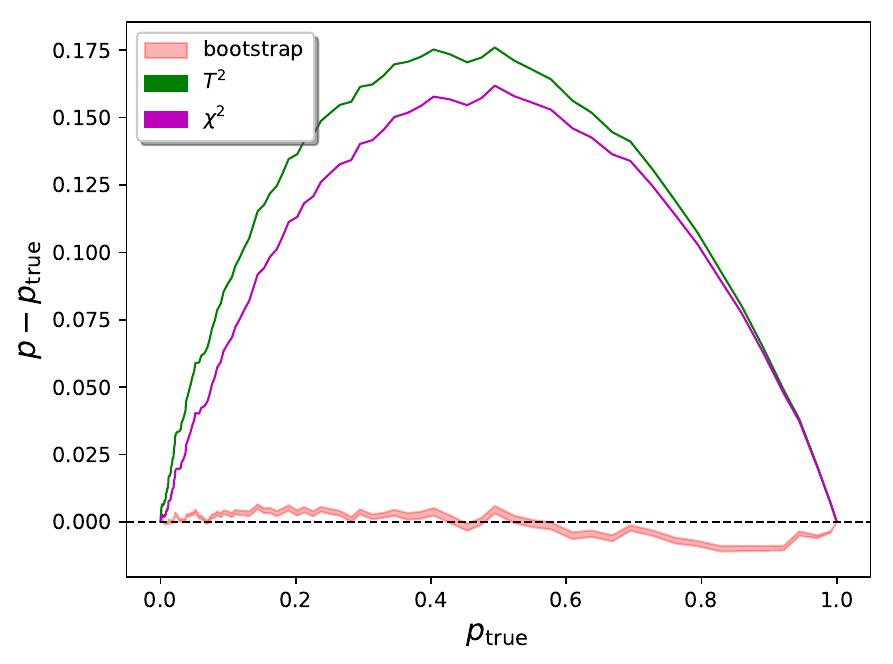}\includegraphics[width=0.49\textwidth]{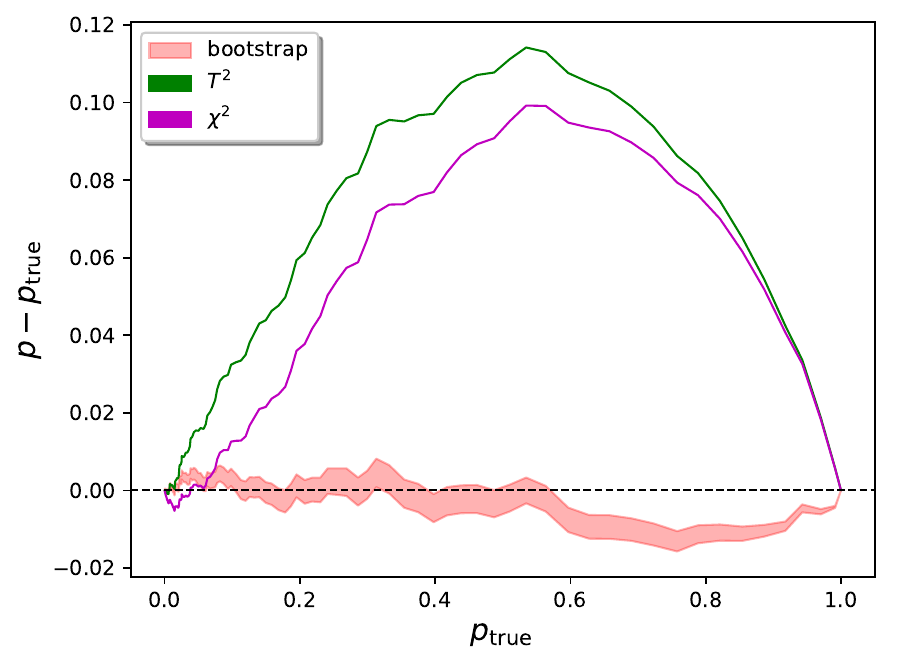}
\caption{The difference between the estimated and true p-values for a fit using the cutoff method (left) and MCM (right), with eigenvalue cutoffs of $\epsilon=1\times 10^{-4}$ and $\bar\epsilon = 0.015$ and the data described in the caption of Fig.~\ref{fig-covmat-evals-mixleft}. The errors on the bootstrap result are obtained from the fluctuations over 30 original ensembles.\label{fig-cutoff-MCM-pdiff-mixleft}  }
\end{figure}

We can test our bootstrap p-value strategy for a toy dataset with a non-trivial eigenspectrum for both the covariance and correlation matrix by introducing a correlation between timeslices. One way to achieve this is to generate data as
\begin{dmath}
C_i(t) = \alpha_t C_i(t-1) + (1-\alpha_t)D_i(t)\label{eq-mixleft-datagen}
\end{dmath}
where $D_i(t){\sim} {\cal N}(\mu,\sigma)$ and $0\leq\alpha_t\leq 1$ ($\alpha_1=0$) induces a timeslice-dependent mixing with the previous timeslice. Choosing $\alpha_t > 0$ on a single timeslice, say $t=T$, induces a low mode whose eigenvalue decreases as $\alpha_{T}$ approaches unity, resulting in a singular covariance matrix at $\alpha_{T}=1$. In Figs.~\ref{fig-covmat-evals-mixleft} and~\ref{fig-corrmat-evals-mixleft} we show an example histogram of the eigenvalue spectrum for estimates of the covariance matrix and correlation matrix, respectively, obtained from many independent ensembles, as well as the corresponding distributions obtained by bootstrapping a single ensemble. We note that the bootstrap eigenspectrum, particularly of the covariance matrix, is quite distorted relative to the true spectrum as a result of finite-$N$ bias, but the well-separated low eigenmodes appear adequately described and we can choose a cutoff that applies consistently to the true and bootstrap distributions. In Fig.~\ref{fig-cutoff-MCM-pdiff-mixleft} we compare the true p-value to the estimated p-value for the bootstrap approach as well as the $T^2$ and $\chi^2$ distributions. We observe that the bootstrap approach reliably estimates the p-value to within errors of a few percent, whereas ignoring the manipulation of the eigenspectrum and using one of the analytic null distributions results in significant overestimates of the p-value.

We conclude that the bootstrap method offers a reliable procedure for obtaining a goodness-of-fit metric for eigenvalue truncation methods, providing the eigenvalues removed are sufficiently separated from the bulk and well resolved.

\subsection{Non-null-consistent data}

In the toy demonstrations above, we have been slightly cavalier in our treatment of the null distribution, because, as noted in Sec.~\ref{sec-goodness-of-fit}, this distribution depends in general not just on the model function but also on details of how null-consistent data fluctuate about this model function, including how they are correlated and how those correlations are treated in obtaining $q^2$. This is clearly true for uncorrelated fits, where a trivial orthogonal rotation among the $T$ time slice data values (under which a correlated fit would be invariant) completely changes the character of the diagonal elements of the covariance matrix. One would expect the eigenvalue truncation methods would likewise suffer. Indeed, the fact that the structural form of the covariance matrix or the model function do not matter for fully-correlated fits to normally-distributed data -- those for which the $T^2$ distribution, which depends only on $N$ and $K$, is a universal description -- is a very special case related to the fact that, for Gaussian random numbers, the distribution of the mean of an ensemble of $N$ samples about the population central value entering the numerator of $q^2$, and that of the differences between individual samples and the ensemble mean (i.e. the covariance matrix) entering the denominator of $q^2$, are statistically independent. (See for example Section 3.4 of Ref.~\cite{GVK02434995X}.) In general therefore, the null distribution is a function not just of $N$ and $K$ but also of the model function, its parameters, and the underlying statistical distribution of the samples in an ensemble (and of course the procedure by which one obtains the covariance matrix). 

The above is largely philosophical for the aforementioned toy demonstrations, as we explicitly asserted that the data are null consistent. By mapping out the distribution of $q^2$ for many such ensembles, either using truly independent ensembles or via the bootstrap, we therefore obtain the appropriate null distribution by definition. The fact that we may have obtained a different distribution for data generated in a different way is irrelevant to the discussion. 

This is not the case when we are uncertain whether the model function is a good description of the data and we are using the p-value for a hypothesis test. There we must be more explicit as to the inputs of the null distribution associated with our hypothesis; specifically, it is not sufficient simply to specify a model function, rather we must also specify the parameters of that function and a model for the fluctuations of the data. Different choices of these inputs define different hypotheses and may result in different null distributions, and hence p-values, for the analysis of the data.  In this section we study the degree to which the proposed bootstrap p-value can be used as a hypothesis test when only the empirical distribution of the data is known.

\subsubsection{Description of the approach}

We begin with a specific model function that we wish to test. While the choice of values for its parameters are entirely arbitrary for defining a hypothesis, a natural choice when performing a hypothesis test is the set obtained from a least-squares fit to our data, as, if the underlying population is truly described by our model function, then those parameters are guaranteed to be close to those of the population and we will be describing the correct null population (up to statistical fluctuations on the parameters) and hence can expect a good and meaningful p-value.

There are numerous ways one might imagine to model the fluctuations of a dataset conforming to a particular hypothesis. However, our desire is to provide a relatively straightforward and data-driven approach that can be applied to a wide variety of analyses, including those of autocorrelated lattice data (which we discuss in the following section). As mentioned above, we must consider two classes of fluctuation: those of the ensemble means about the population means, and those of the samples within an ensemble about the ensemble mean. For non-Gaussian data, these two fluctuations are not statistically independent, hence the most straightforward way to model them is to generate many representative ensembles of fake data. If we restrict ourselves to hypotheses in which the fluctuations match those of our data, bootstrap resampling allows us to generate such ensembles without any knowledge or assumptions on the underlying distribution of the data. 

The fake data ensembles we use to model the fluctuations must of course be centered on our model with the parameters chosen as above, in order to also be null consistent with our hypothesis for the population means. If we are employing the bootstrap to model the fluctuations, appropriate resampled ensembles can be generated by the recentering method described in Eq.~\ref{eq-recentering}, which we formerly employed to correct for the statistical fluctuations of the original ensemble. 

In addition to the above, we must continue to treat these data as we do in our standard analysis procedure, fitting each one independently so as to capture the correct number of degrees of freedom in our null distribution. 

Together, we find that the bootstrap p-value estimation procedure described in Sec.~\ref{sec-bootstrap-approach-descr} can be applied entirely without modification in the context of hypothesis testing, at least for a particular class of hypotheses.

\subsubsection{Internal consistency of the null hypothesis}

The model of the fluctuations described above does not take into account any changes in their distribution that might be associated with the choice of model function. At first glance these two quantities may appear independent: The estimate of the covariance matrix obtained from our ensemble via Eq.~\ref{eq-covmatdef} is invariant under arbitrary, possibly timeslice-dependent, shifts in the samples comprising an ensemble: $C_i(t) \to C_i(t) + f(t)$. Furthermore, we assume that $N$ is sufficiently large that the central limit theorem applies and the ensemble means are normally distributed, implying that the fluctuations of the ensemble means about the population means are also independent of the values of those means (and hence the model function). However these facts do not shield us from contradictions between the model function and the underlying probability distribution governing the fluctuations of the data. We gave a concrete example of this in our Overview, Sec.~\ref{sec-overview}, for lattice data describing the two-point correlation function for a system of two identical particles that is described by a single energy eigenstate. The true functional form for this system contains a single exponential term and a constant term associated with around-the-world propagation of a single particle, but we consider a hypothesis where this constant term is excluded. Regardless of our choice of model function, the fluctuations that we capture in our data-driven model are sensitive to the presence of this constant term, which may significantly alter the eigenspectrum of the covariance matrix (possibly even making the matrix singular), and we are left with a contradiction between our two models. Thus, while we retain the capability of reliably testing a specific hypothesis, that hypothesis may not be internally consistent.

This potential contradiction does not necessarily spell doom for our approach, as our primary goal is to perform a hypothesis test for the model function and its parameters, which have a physical interpretation, and we would expect such a test to be much more sensitive to an incorrect choice of model function or to the gross effects of manipulating the form of the covariance matrix, than to a subtlety of the fluctuations of this matrix. Indeed, regardless of their form, the fluctuations of the covariance matrix fall as $1/N$ whereas those of the ensemble means (which are normally-distributed) fall as $1/N^{1/2}$. We can also be assured that if we do happen to choose the right model function and its parameters, then our dataset will be null consistent with the distribution we obtain and we are likely to observe a good p-value. Together, this suggests that a hypothesis test based on the p-value we obtain remains a good discriminator for a poor choice of model function, and, especially for cases involving an alternative definition of the covariance matrix, is certainly a more reliable goodness-of-fit metric than that obtained by assuming the $T^2$ or $\chi^2$ distribution, or of simply abandoning estimation of this quantity altogether.

\subsubsection{Toy model demonstration}

\begin{figure}[t]
\centering
\includegraphics[width=0.49\textwidth]{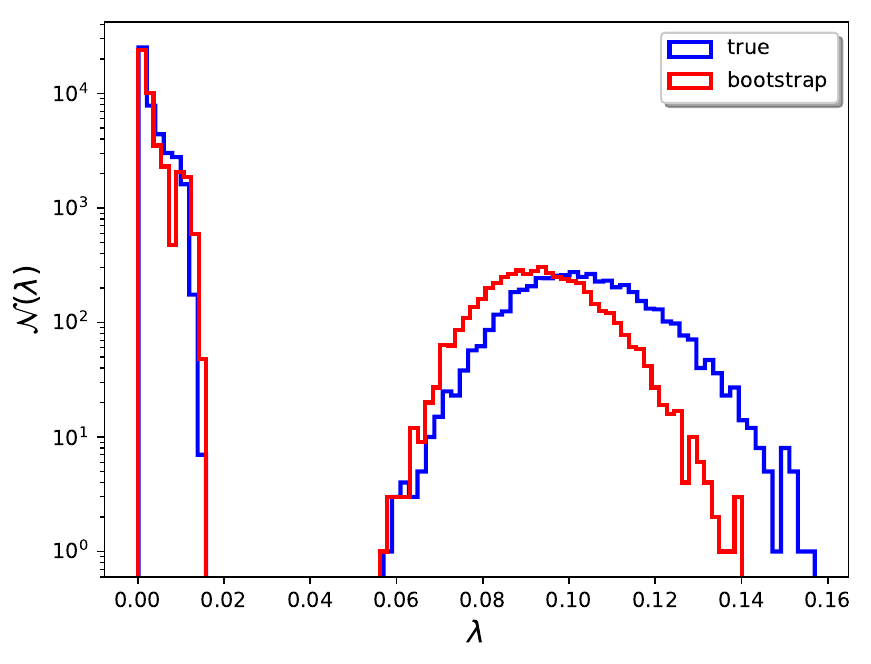}
\caption{A histogram of the complete eigenspectrum for the estimated covariance matrix from ensembles generated as described in Eq.~\ref{eq-toy-data-constshift} with the parameters specified in the text. Results are shown for 5000 independently-generated ensembles (``true") and as many bootstrap ensembles (``bootstrap") generated from a fixed original ensemble. \label{fig-covmat-evals-exp-plus-const}  }
\end{figure}

In this section we demonstrate the bootstrap p-value technique in the context of a hypothesis test based on the scenario described above. To this end, we will employ toy data generated with $N=100$, $T=10$ and the following temporal dependence:
\begin{dmath}
C(t) = A e^{-mt} + c \label{eq-toy-data-constshift}
\end{dmath}
where $m_{\rm in}=0.2$, and the coefficient of the exponential term is sampled independently on each timeslice as $A_{\rm in}\sim {\cal N}(1.0,1.0)$, corresponding to a 10\% error-on-the-mean for this quantity and ensemble size. Here and below, the subscript ``in" denotes the input values to our data generation, as opposed to those obtained from a fit or otherwise. The constant $c_{\rm in}\sim {\cal N}(1.0,1.0)$ is chosen to be common to all timeslices and introduces a large and strongly fluctuating eigenvalue of the covariance matrix that has a marked impact upon the eigenspectrum, as shown in Fig.~\ref{fig-covmat-evals-exp-plus-const}. In this figure we also compare the bootstrap and true eigenspectra, finding reasonable agreement.

\begin{figure}[t]
\centering
\includegraphics[width=0.49\textwidth]{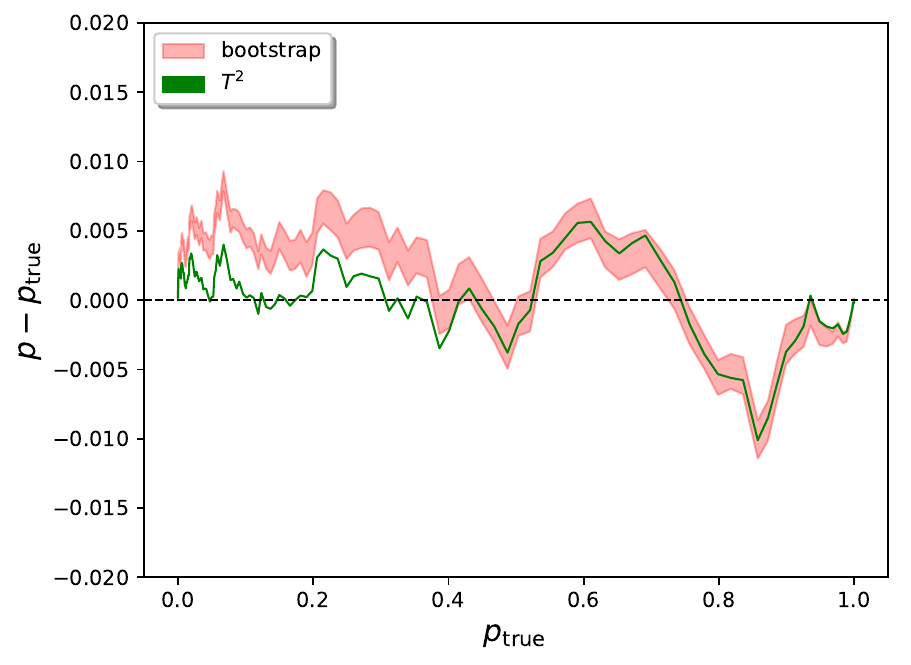}\includegraphics[width=0.49\textwidth]{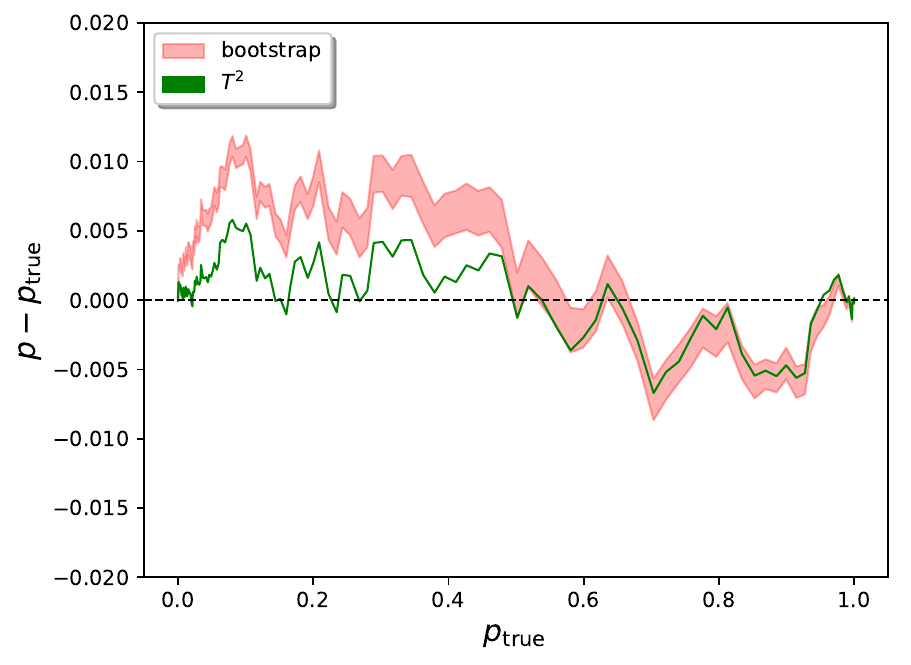}

\caption{The difference between the estimated and true p-values for the null distribution of correlated fits to data generated as described in Eq.~\ref{eq-toy-data-constshift} with the parameters specified in the text, using a model function that includes the constant as a free parameter (left) and one that does not (right). In both cases, the fluctuations are modeled based on those of the data. The errors on the bootstrap result are obtained from the variation over 30 original ensembles. Also shown is the same comparison for the Hotelling $T^2$ distribution with $N=100$, $K=7$ (left) and $N=100$, $K=8$ (right). The lack of smoothness in the $T^2$ curves when viewed at this level of precision results from the true p-value to which it is compared being estimated from a finite number (5000) of independent ensembles. The bootstrap results also likewise fluctuate due to being estimated from 5000 bootstrap resampled ensembles. \label{fig-pdiff-corr-exp-plus-const}  }
\end{figure}

\begin{figure}[t]
\centering
\includegraphics[width=0.49\textwidth]{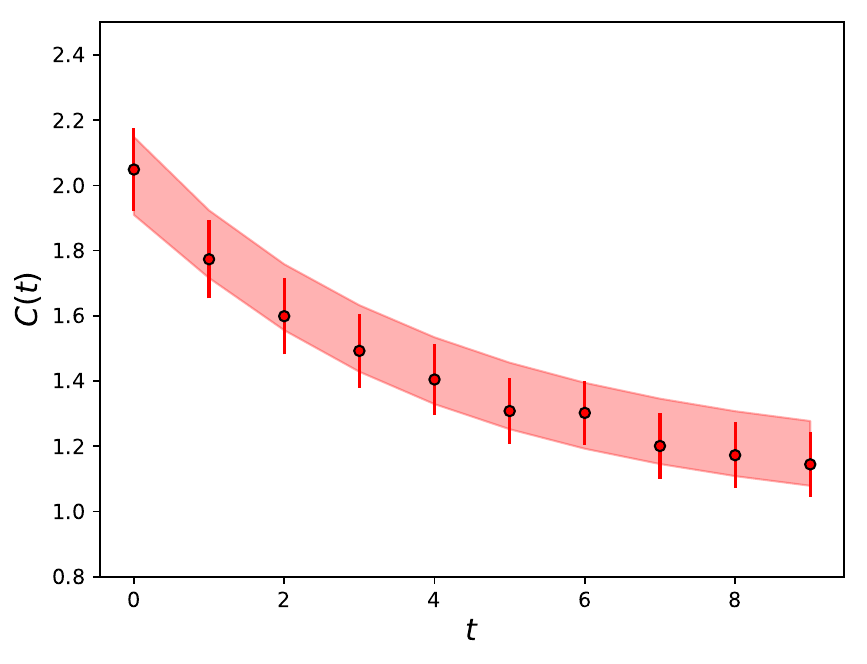}\includegraphics[width=0.49\textwidth]{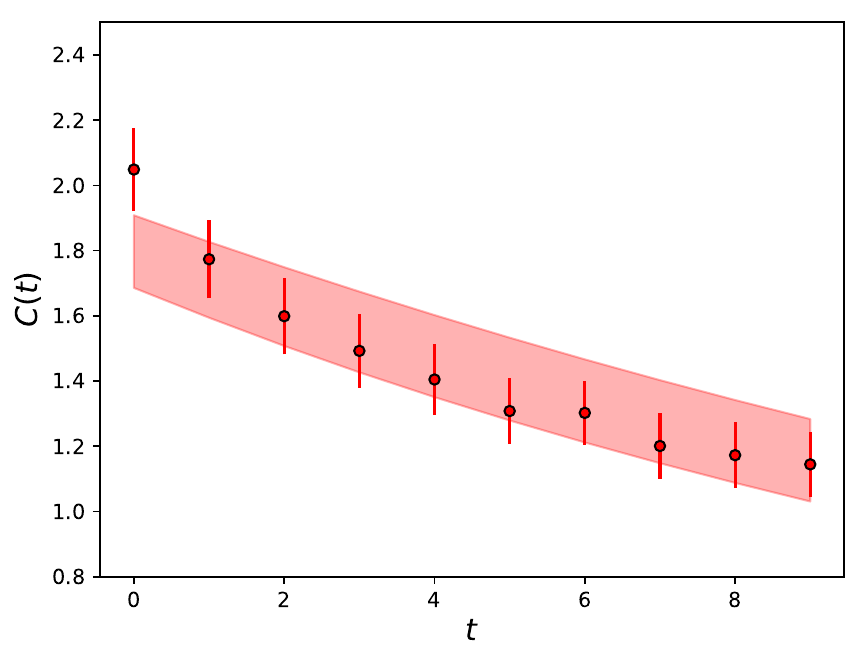}

\caption{A plot of data generated as described in Eq.~\ref{eq-toy-data-constshift} with the parameters specified in the text, compared to the fit curve obtained using a model function that includes the constant as a free parameter (left) and one that does not (right). The parameters of the fit that includes the constant term are $A_{\rm fit}=0.957(99)$, $m_{\rm fit}=0.245(54)$, $c_{\rm fit}=1.07(14)$, and for that without the constant term, $A_{\rm fit}=1.794(111)$, $m_{\rm fit}=0.049(17)$. The errors are obtained via bootstrap resampling. \label{fig-fit-vs-data-exp-plus-const} }
\end{figure}

\begin{table}[tb]
\begin{tabular}{cc|ccc}
\hline\hline
Fit type & Constant term & $p_{\rm true}$ & $p_{\rm boot}$ & $p_{\rm T^2}$    \\
\hline
Correlated & Y & 0.932 &  0.934(5)(4) & 0.931 \\
           & N & 0.052 &  0.060(3)(3) & 0.055 \\
\hline
Uncorrelated & Y & 0.960 & 0.941(13)(25) & 1.000 \\
             & N & 0.053 & 0.035(25)(18) & 0.928
\end{tabular}

\caption{A table of the p-values obtained via the true ($p_{\rm true}$), bootstrap ($p_{\rm boot}$) and Hotelling $T^2$ ($p_{\rm T^2}$) null distributions for correlated and uncorrelated fits to data generated as described in Eq.~\ref{eq-toy-data-constshift} with the parameters specified in the text. The notation Y and N in the second column indicates that the constant was included, and was not included, as a free parameter, respectively. The two errors quoted for the bootstrap result were obtained with 20 independent original ensembles, and 20 bootstrap resampled ensembles, respectively, following the procedure given in Sec.~\ref{sec-p-value-stat-err}. \label{tab-fit-pvals-corr-uncorr-exp-plus-const} }

\end{table}

We first demonstrate that performing a correlated fit to such data with a model function that includes the constant as a free parameter (i.e. using Eq.~\ref{eq-toy-data-constshift} as the model function) results in a good fit. The original ensemble is known to be null-consistent in this case, and we can simply apply the bootstrap method as we did in Sec.~\ref{sec-toy-model-demo-null-consistent} without needing to take care defining the hypothesis. As the two terms on the right-hand side of Eq.~\ref{eq-toy-data-constshift} are independent and normally distributed their sum is also normally distributed, hence the true null distribution is $T^2$ with $N=100$ and $K=7$. In Fig.~\ref{fig-pdiff-corr-exp-plus-const} we demonstrate that the bootstrap method indeed reproduces the true and $T^2$ distributions. The corresponding p-values for our single original ensemble are given in Tab.~\ref{tab-fit-pvals-corr-uncorr-exp-plus-const}, where we observe an excellent fit as expected. A plot of the data compared to the fit is shown in Fig.~\ref{fig-fit-vs-data-exp-plus-const}, with the fit parameters given in the caption.

We now perform a different fit to the same ensemble, in which we fix (freeze) the constant term to $c=0$, thus disagreeing with the true functional form for the population means. The corresponding fit curve and its parameters are also provided in Fig.~\ref{fig-fit-vs-data-exp-plus-const}, where we observe a much lower fitted mass than the true value, and a much larger coefficient. The hypothesis that we wish to test is for this (incorrect) model function and the above fit parameters, and uses the data to model the fluctuations. We expect this test to fail, despite those parameters being optimized to best describe the data, as it is unlikely that such data as shown in Fig.~\ref{fig-fit-vs-data-exp-plus-const}, could be generated by statistical fluctuations about the model function, particularly at small-$t$. We will label the parameters obtained from this fit with a subscript ``fit", i.e. $A_{\rm fit}$, $m_{\rm fit}$. We will also write the frozen constant term as $c_{\rm frz}$($=0$) so as to make the discussion below more general. The complete model for the temporal dependence of the ensemble means is thus
\begin{dmath}
M(t) = A_{\rm fit} e^{-m_{\rm fit}t} + c_{\rm frz}\,.
\end{dmath}
To perform the test we require the null distribution for the hypothesis specified above. As we continue to perform a correlated fit to normal data, $T^2$ remains the appropriate null distribution, albeit with $K=8$ to account for the reduced number of free parameters. We will attempt to estimate this distribution using the bootstrap procedure described above. As before, we also wish to compare the bootstrap null distribution to a ``true" null distribution with the same data-driven hypothesis for the fluctuations. This can be achieved by generating independent ensembles according to Eq.~\ref{eq-toy-data-constshift}, i.e. for which $c$ fluctuates randomly around unity, and then recentering those data to match our model function:
\begin{dmath}
C_i(t) \to C_i(t) - \trueavg{C(t)} + M(t)\,,
\end{dmath}
where the population means are
\begin{dmath}
\trueavg{C(t)} = \bar A_{\rm in} e^{-m_{\rm in}t} + \bar c_{\rm in}\,,
\end{dmath}
with $\bar A_{\rm in}=\bar c_{\rm in} = 1.0$ (and $m_{\rm in}=0.2$), such that $\bar C(t) \to M(t)$. (Again this recentering leaves the covariance matrix unaffected.) The recentering employed for the bootstrap ensembles is that described in Sec.~\ref{sec-bootstrap-approach-descr}, which differs from the above only in replacing the population average $\trueavg{C(t)}$ with the corresponding mean of the empirical distribution:
\begin{dmath}
{ C^{*,b}_i(t) \to C^{*,b}_i(t) - \trueavg{C(t)}_R + M(t) = C^{*,b}_i(t) - \bar C(t) + M(t) }
\end{dmath}
We can now use the independent and bootstrap recentered ensembles to generate the ``true" and bootstrap null distributions, respectively. Of course our data-driven hypothesis for the fluctuations is now (deliberately) in conflict with our model function due to the large eigenvalues brought about by the constant shift; however, this is irrelevant in this particular case as described above, and we again observe excellent agreement between the bootstrap, true and $T^2$ distributions in Fig.~\ref{fig-pdiff-corr-exp-plus-const}. The p-value is also given in Tab.~\ref{tab-fit-pvals-corr-uncorr-exp-plus-const}, where we find a poor fit as expected.

\begin{figure}[t]
\centering
\includegraphics[width=0.49\textwidth]{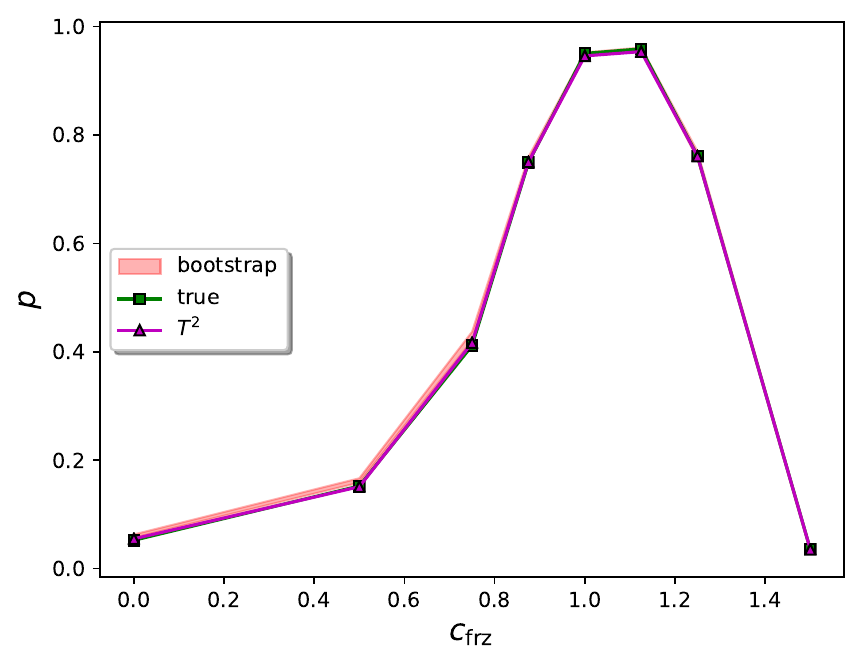}\includegraphics[width=0.49\textwidth]{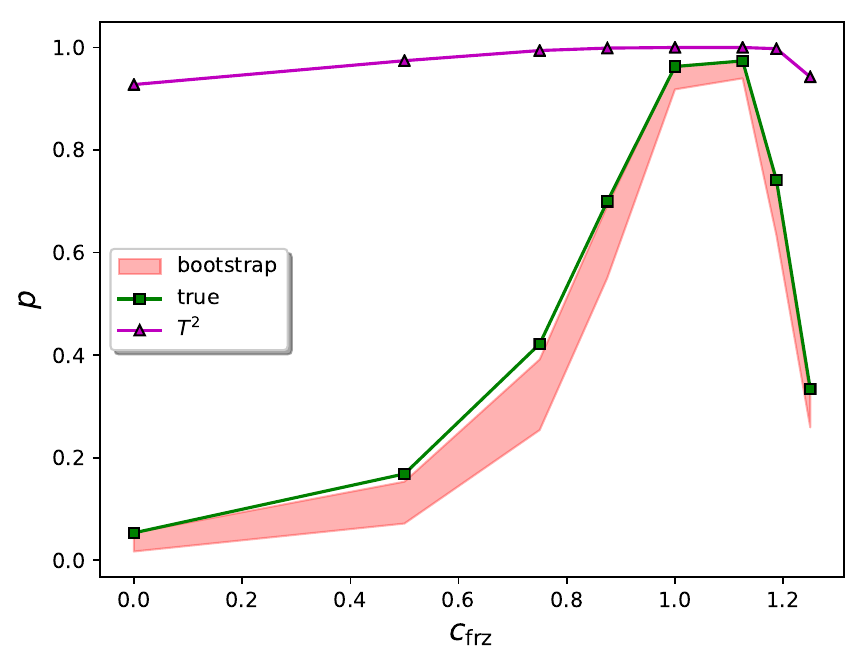}

\caption{The p-value as a function of the frozen constant fit parameter $c_{\rm frz}$, for data generated as described in Eq.~\ref{eq-toy-data-constshift} and correlated fits (left) and uncorrelated fits (right). In both cases we compare the bootstrap and true p-values vs those of the Hotelling $T^2$ distribution, which is the correct analytic description for the correlated fits but it not appropriate for the uncorrelated fits. \label{fig-p-vs-frozen-const} }
\end{figure}

Continuing with these frozen fits, we expect that we will find a better p-value as $c_{\rm frz}$ is brought closer to the true population value. In Fig.~\ref{fig-p-vs-frozen-const} we plot the p-value as function of $c_{\rm frz}$, where we indeed observe the p-value peaking for $c_{\rm frz}$ around the population mean value of unity. Taken together, we observe all the expected characteristics of a sensible hypothesis test of our model function, albeit for a case where the null distribution is independent of the subtleties of the underlying sample distribution.

We can extend this demonstration to a case where those subtleties are important by switching to uncorrelated fits. Here we cannot compare to a known analytic form, and our hypothesis is only internally consistent when the model function contains an appropriate constant term; nevertheless, we still expect that our approach will evince a poor fit when fitting without the constant term and a good fit when it is included as a free parameter. (In this latter case the internal contradiction in the hypothesis will of course disappear.) In Tab.~\ref{tab-fit-pvals-corr-uncorr-exp-plus-const} we list the true and bootstrap p-values for both cases, where we observe the expected behavior. We also expect that frozen fits with $c_{\rm frz}$ approaching the population value will exhibit similar behavior to that seen for correlated fits above; we observe in Fig.~\ref{fig-p-vs-frozen-const} that this is indeed the case, with the p-value peaking around the population value of the constant. 

In this plot we also show the p-values obtained assuming the $T^2$ distribution, which dramatically fail to provide a sensible hypothesis test, with p-values all in the neighborhood of unity even if the constant term is frozen to zero.

\section{Working with autocorrelated data}
\label{sec-autocorrelated-data}

The bootstrap procedure described above remains usable even in the context of autocorrelated data. For example, if the data are binned to a sufficient degree that the binned samples are effectively independent and the number of binned samples is sufficiently large, the bootstrapping can be applied to the binned samples without modification. However, if binning is not used, it is necessary to define an appropriate bootstrap resampling strategy that preserves as best as possible the underlying autocorrelations. One way to achieve this is through the (non-overlapping) ``block bootstrap"~\cite{carlstein_nbb}.

\subsection{Non-overlapping block bootstrap}

The non-overlapping block bootstrap (NBB) method is described as follows: For a block size $B$, divide the ensemble of size $N$ into $N/B$ equal-sized subsets (``blocks") of consecutive points (truncating any additional data if necessary), and label each of these blocks with an index $b\in\{1..N/B\}$. For each resampled ensemble, draw $N/B$ uniformly-distributed random integers in the range ${1..N/B}$ and lay out the blocks of $B$ points indexed by those values consecutively to form an array of size $N$. 

The minimum separation between two effectively independent samples might be estimated as $2\tau_{\rm int}$, where $\tau_{\rm int}$ is the integrated autocorrelation time,
\begin{dmath}
\tau_{\rm int} = \frac{1}{2} + \sum_{\Delta=1}^{\infty} \rho(\Delta)\,,
\end{dmath}
and $\rho(\Delta)$ is the autocorrelation function for samples separated by a distance $\Delta$ within the ensemble,
\begin{dmath}
\rho(\Delta) = \left\langle ( C_i - \overline{C} )( C_{i+\Delta} - \overline{C} )\right\rangle / \sigma^2
\end{dmath}
for some quantity $C$ of variance $\sigma^2$. Thus, providing $2\tau_{\rm int}\ll B$, the block encapsulates the majority of the autocorrelations for the samples it contains. This restriction is the same for binning; in fact, for the ensemble means, binning the data over bins of size $B$ then performing a regular bootstrap resampling produces identical results to the NBB with block size $B$. In both cases, the reduction of the number of effective samples from $N$ to $N/B$ enhances the bias in the bootstrap procedure, as we will discuss further below. 

The benefit of using the NBB over binning is that one obtains resampled ensembles of the same size as the original ensemble, and which, up to biases, can be treated for the purposes of analysis as independent ensembles of Monte Carlo data, generated with the same degree of autocorrelation. In the following section we describe one way in which this can be exploited.

\subsection{Autocorrelation-ignoring covariance matrix}
\label{sec-autocorr-ignoring-covmat}
In the RBC \& UKQCD collaboration's analysis of isospin $I=0$ two-pion correlation functions we encountered an issue, described in detail in Ref.~\cite{Kelly:2019wfj}, that ultimately led to the developments described in this document. The simulation comprised 741 measurements separated by 4 molecular dynamics time units (MDTU), roughly equal to the measured integrated autocorrelation time. We binned the data in order to account for any residual autocorrelation effects on the statistical error, increasing the bin size to find the plateau in the measured error. For the kaon and pion two-point functions, we found this to occur at bin sizes $B\gtrsim 6$ measurements (24 MDTU); however, for the two-pion correlation functions no such plateau was observed, rather the error continued to grow with increasing rate as the bin size was increased. 

The key difference between the pion/kaon data and the two-pion data is the size of the covariance matrix: For the former, the number of data points included in the fit was ${\cal O}(15-20)$; however, the two-pion data comprised correlators with 3 different source/sink operators and multiple timeslices, totaling 66 distinct data points for our best fit. The corresponding covariance matrix is therefore very large: $66\times 66$. 

\begin{figure}[tb]
\captionsetup[subfigure]{justification=centering}
\centering
\begin{subfigure}{0.49\textwidth}
\includegraphics[width=\textwidth]{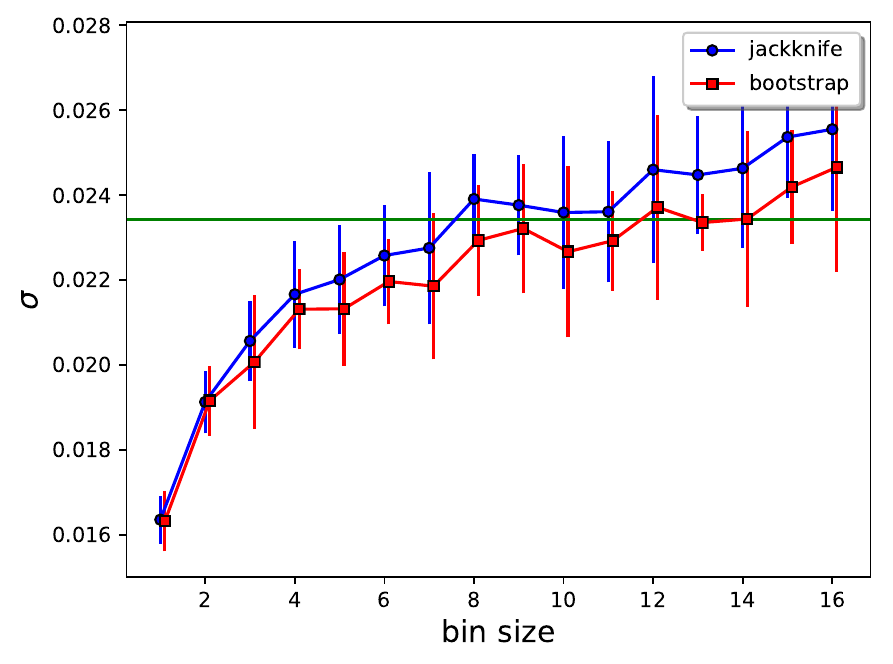}
\caption{$T=5$}
\end{subfigure}
\hfill
\begin{subfigure}{0.49\textwidth}
\includegraphics[width=\textwidth]{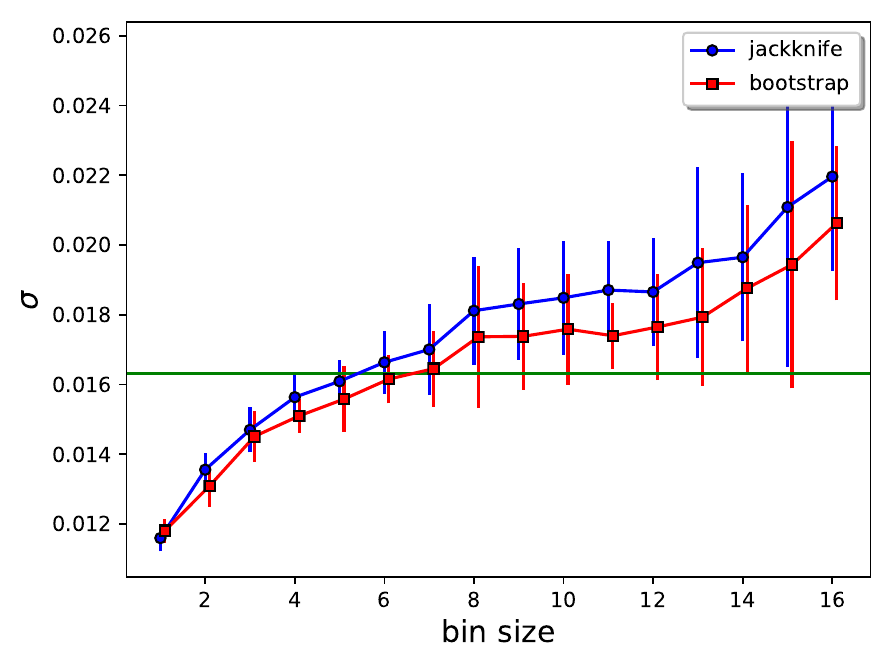}
\caption{$T=10$}
\end{subfigure}
\hfill
\begin{subfigure}{0.49\textwidth}
\includegraphics[width=\textwidth]{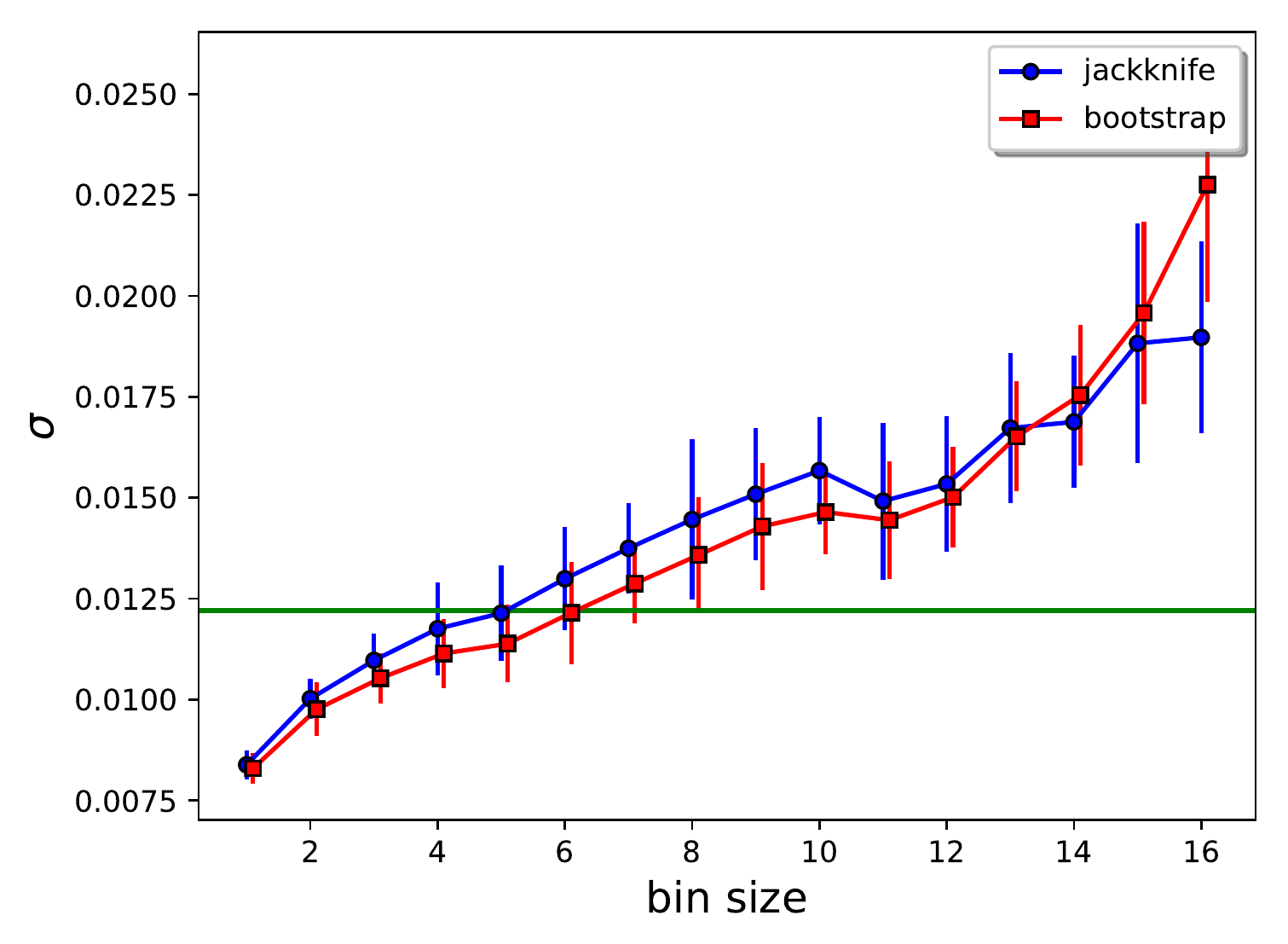}
\caption{$T=20$}
\end{subfigure}
\hfill
\begin{subfigure}{0.49\textwidth}
\includegraphics[width=\textwidth]{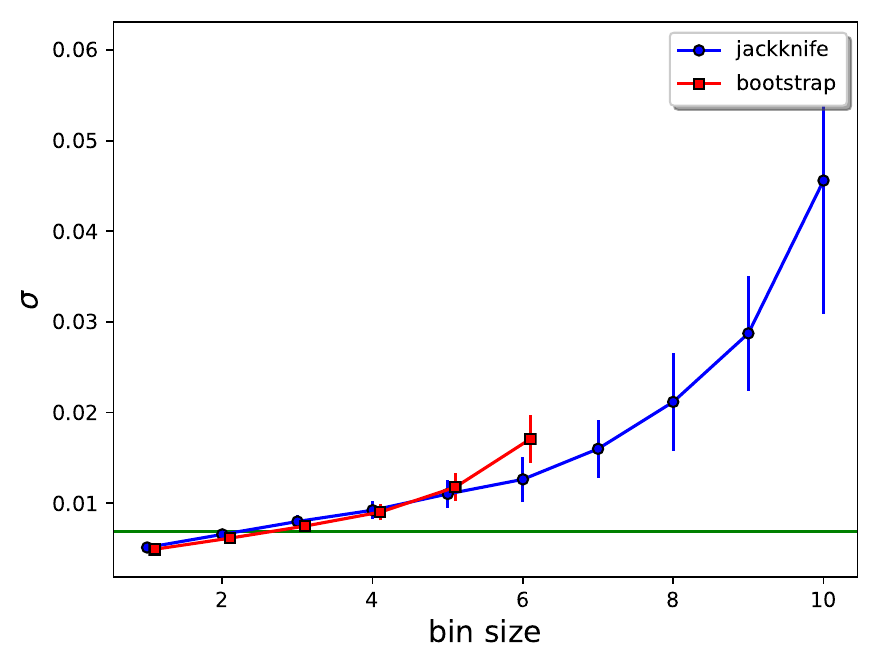}
\caption{$T=65$}
\end{subfigure}
\hfill
\caption{The bootstrap- and jackknife-estimated statistical error on the result of a constant fit to $T$ timeslices of independent data, as a function of bin size, for several values of $T$. Data for each timeslice are generated from a normal distribution ${\cal N}(0,1)$ using the Metropolis algorithm tuned to obtain an 84\% acceptance and an integrated autocorrelation time of $\tau_{\rm int}\approx 12$ Metropolis steps. (Note our conventions are such that this value for $\tau_{\rm int}$ indicates $2\tau_{\rm int}\approx 2$ separation between independent samples.) We employ 750 samples with 12 Metropolis steps between each sample. The error bars show the variation in the error over 10 different ensembles. The horizontal green line shows the true error on the fit result obtained by repeating the analysis with 500 independent original ensembles. \label{fig-binerr-metropolis} }
\end{figure}

\begin{figure}[tb]
\captionsetup[subfigure]{justification=centering}
\centering
\begin{subfigure}{0.49\textwidth}
\includegraphics[width=\textwidth]{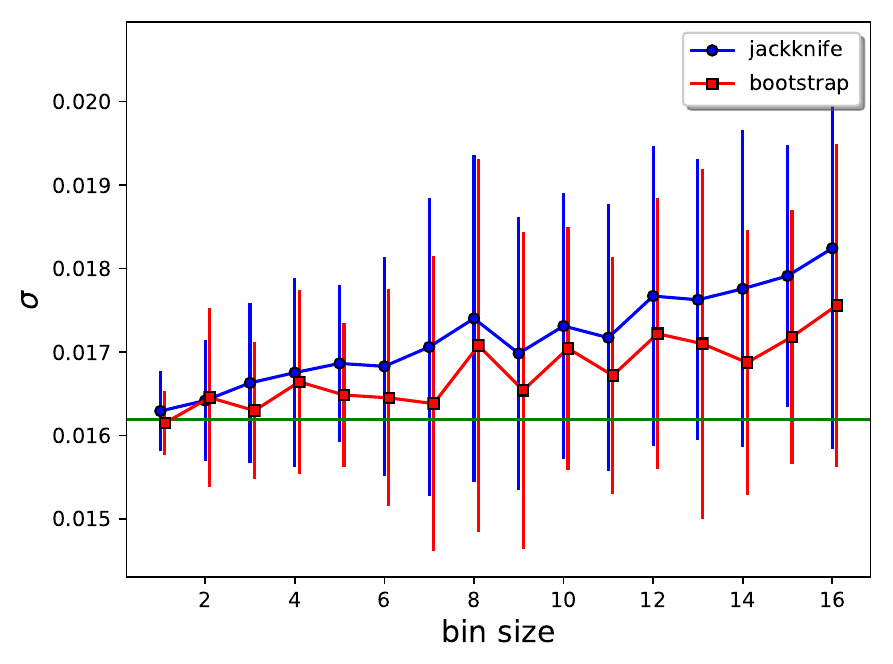}
\caption{$T=5$}
\end{subfigure}
\hfill
\begin{subfigure}{0.49\textwidth}
\includegraphics[width=\textwidth]{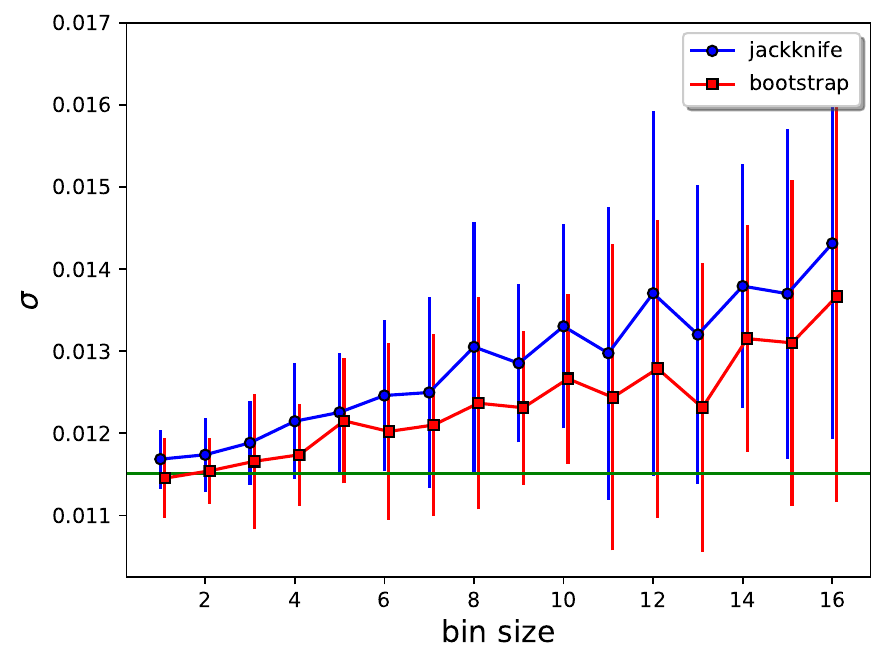}
\caption{$T=10$}
\end{subfigure}
\hfill
\begin{subfigure}{0.49\textwidth}
\includegraphics[width=\textwidth]{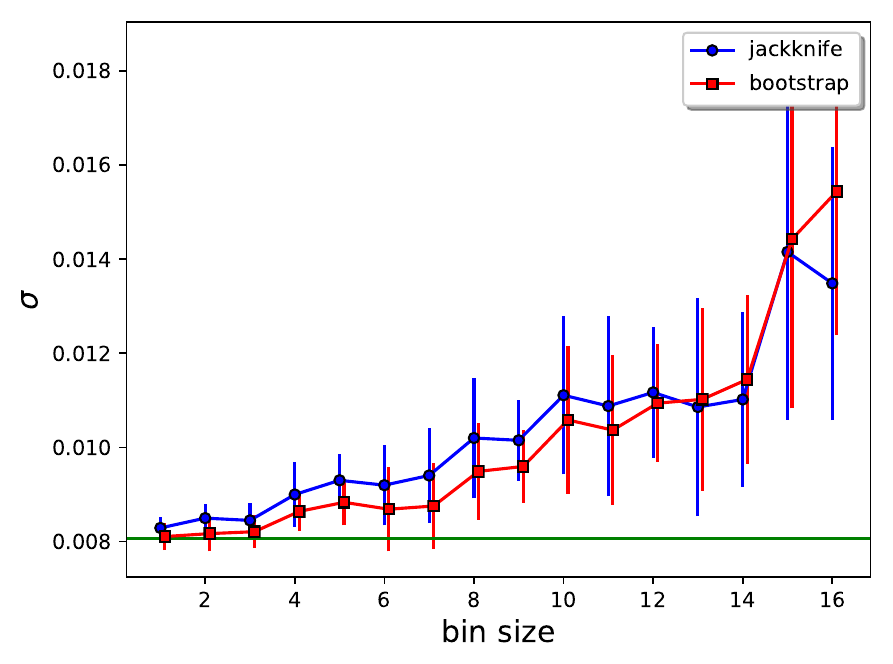}
\caption{$T=20$}
\end{subfigure}
\hfill
\begin{subfigure}{0.49\textwidth}
\includegraphics[width=\textwidth]{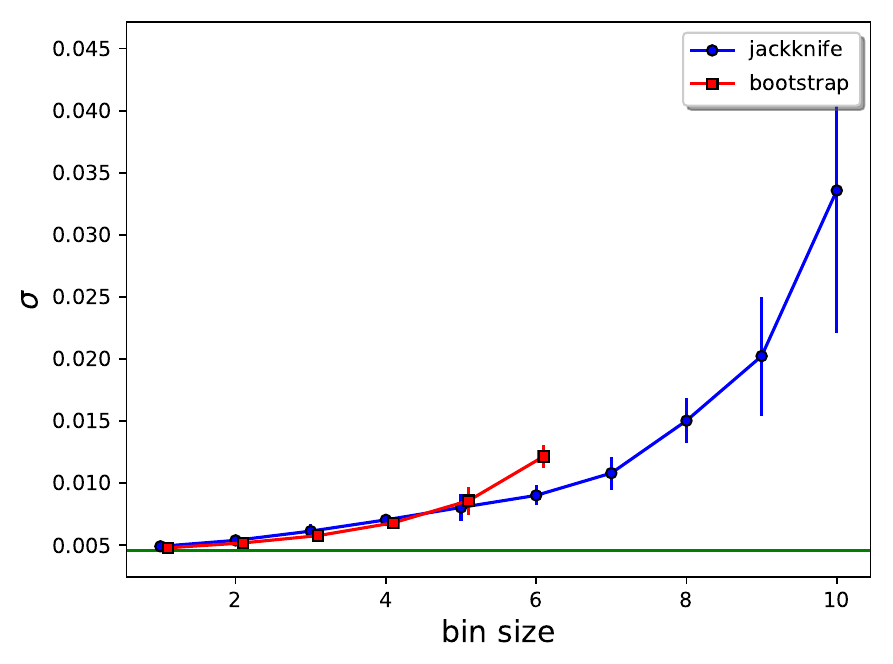}
\caption{$T=65$}
\end{subfigure}
\hfill
\caption{A repeat of the analysis shown in Fig.~\ref{fig-binerr-metropolis} but for independent, non-autocorrelated samples. \label{fig-binerr} }
\end{figure}

We can reproduce the behavior described above with our simple toy model of unit-variance Gaussian data with constant time dependence, for which the population covariance matrix is diagonal with $T$ eigenvalues all equal to $1/N$. In Fig.~\ref{fig-binerr-metropolis} we plot the jackknife and bootstrap error estimates as a function of bin size for a constant fit to autocorrelated Gaussian data generated using the Metropolis algorithm. For small numbers of timeslices, $T=5$ and 10, we observe the expected behavior; the statistical error estimate grows as we increase the bin size, before stabilizing at a value consistent with the true statistical error (although there is some evidence of growth at larger bin sizes). However, for larger $T$ we begin to see the error estimates continue to rise significantly beyond where the plateau is expected. For $T=20$ a reasonable error estimate can still be obtained for bin sizes in the 6-10 range, but for $T=65$, where the covariance matrix is comparable in dimension to that of the two-pion fit example, the error growth becomes very rapid at larger bin sizes and we observe the breakdown mode described above. We plot a repeat of this analysis but for independent, non-autocorrelated data in Fig.~\ref{fig-binerr}, in which we observe the same breakdown mode occurring at larger bin sizes, implying it is unrelated to the autocorrelations.

The breakdown of the binning approach can be explained as follows: It is clear that fits to binned data must fail when the dimension of the correlation matrix exceeds the number of samples used to determine it,
\begin{dmath}
\frac{N}{B} \leq T.
\end{dmath}
When this inequality is obeyed the covariance matrix becomes singular. (Recall that the covariance matrix can be viewed as a sum of the outer product of $T$-dimensional vectors, one for each sample.)  As we increase the bin size and $N/B$ approaches $T$, the covariance matrix develops low eigenmodes that will dominate the inverse and thus the determination of the minimum of $q^2$. Exactly how this translates into an increasing error on the fit parameters relates to how much these low modes fluctuate between resampled ensembles, or indeed between different independent ensembles, which of course depends on the distribution of the underlying data. However, it is reasonable to expect that the fluctuations of the low eigenmodes will increase as we approach the singular point. Eventually, these fluctuations dominate the error on the fit result and the procedure becomes non-viable.

Two other phenomena are observable in Figs.~\ref{fig-binerr-metropolis} and~\ref{fig-binerr}, the first is that for larger bin sizes, the bootstrap appears to consistently produce lower error estimates than the jackknife, although within the error-on-the-error. This is a manifestation of the bootstrap bias, and originates from the fact that the variance of a bootstrap resampled ensemble is biased low due to the existence of repetitions of the same value, creating regions within the ensemble with zero variance. Such repetitions become more common as the number of samples diminishes. The second observation is that the bootstrap error estimate appears to break down more rapidly than the jackknife, becoming very challenging in the $T=65$ case for bin sizes larger than 6, due to the appearance of bootstrap resampled ensembles with covariance matrices that are sufficiently ill-conditioned that the numerical inverse fails. This likely relates to the bias in the eigenspectrum observed in Sec.~\ref{sec-eval-trunc-demo}, and also to the fact that jackknife resampled ensembles differ only in a single binned sample from the original ensemble; hence we might expect greater stability assuming the ensemble covariance matrix of the original ensemble is invertible.

\begin{figure}[t]
\centering
\includegraphics[width=0.49\textwidth]{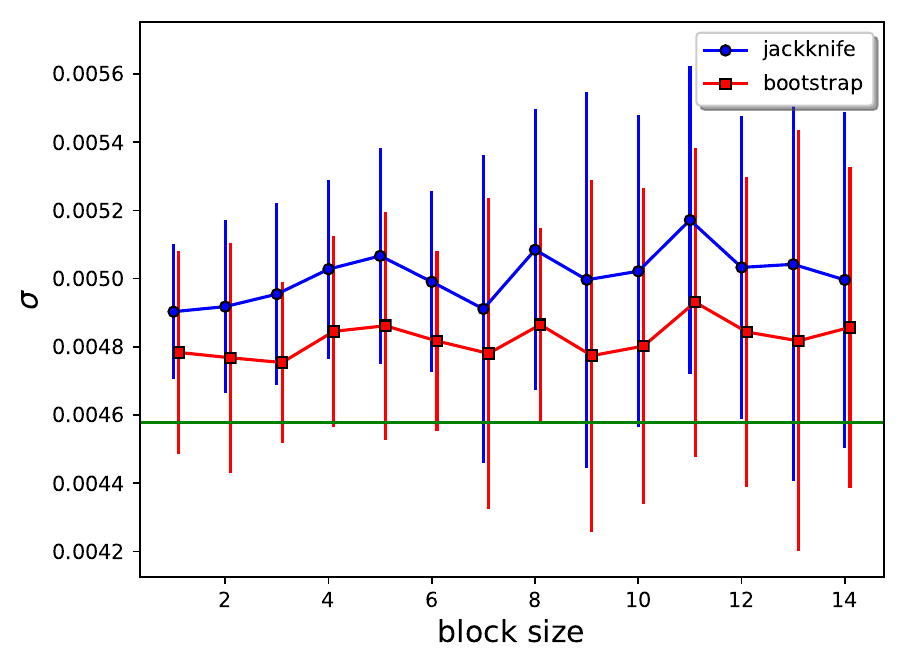}
\hfill
\includegraphics[width=0.49\textwidth]{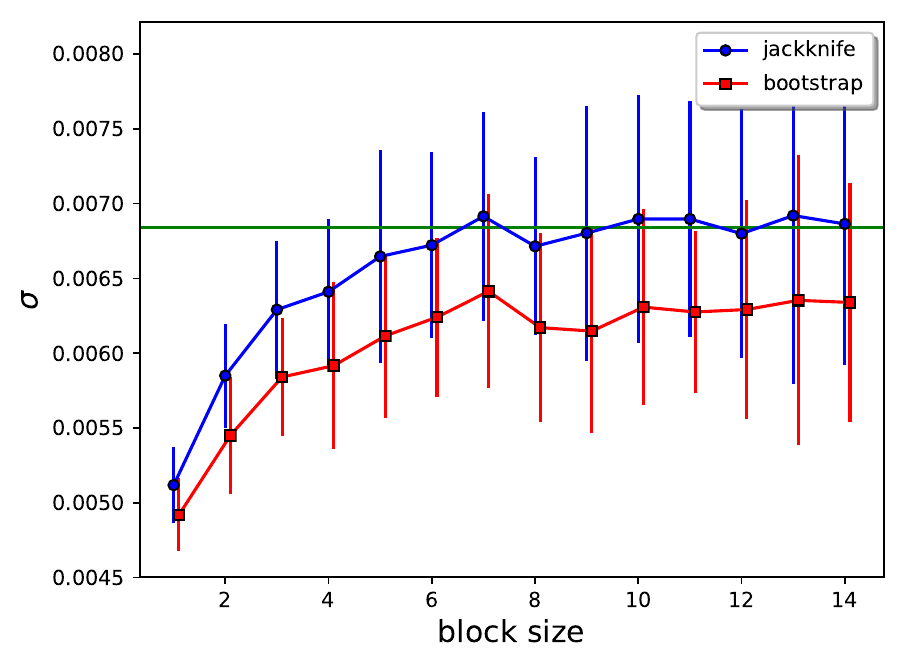}
\caption{The block bootstrap- and jackknife-estimated statistical error on the result of a constant fit to $T=65$ timeslices of data generated as per Fig.~\ref{fig-binerr-metropolis}, as a function of block size. The left plot shows the result for independent data samples and the right for autocorrelated data. The horizontal green line shows the true error on the fit result obtained by repeating the analysis with 500 independent original ensembles.\label{fig-autocorr-avoid}}
\end{figure}

In order to avoid these issues, we adopted the approach of computing the covariance matrix from the original, unbinned data, thereby maintaining a large number of samples with which to obtain a stable determination at the cost of ignoring the effects of autocorrelations. We refer to this as the ``autocorrelation-ignoring covariance matrix" approach. In order to determine the appropriate statistical errors, the NBB approach was employed to generate resampled ensembles that attempt to preserve these autocorrelations and thus provide a good estimate of the fluctuations were we to repeat this procedure on many independent ensembles. In Fig.~\ref{fig-autocorr-avoid} we plot the statistical error estimate on the $T=65$ fit result as a function of block size, using data generated in the same way as above. We plot the error obtained using the NBB procedure, as well as the equivalent ``block jackknife'' procedure in which the resampled ensembles are obtained by elimination of $B$ consecutive data points rather than just one. For the case of independent, non-autocorrelated data we find no significant variation in the estimated error for all block sizes in the range 1-14 that we tested, and that the error estimates are consistent with the true error. For autocorrelated data, we observe the expected rise in the error followed by a plateau at a value again consistent with the true error. Together, this indicates that this procedure resolves the breakdown observed above, and thus allows for reliable statistical error estimation for fits with large covariance matrices and autocorrelated data.

\subsection{Bootstrap p-values for the autocorrelation-ignoring covariance matrix approach}
\label{sec-bootstrap-autocorr-ignore-pvalue}

In neglecting the effect of autocorrelations on the covariance matrix in this "autocorrelation-ignoring" approach, we are subtly modifying the definition of $q^2$ in a way that again invalidates the use of an analytic null distribution to obtain the goodness-of-fit. However, using the bootstrap procedure for determining the null distribution (employing the NBB procedure in the generation of the resampled ensembles), we are able to assign a p-value to our measurements.

\begin{figure}[t]
\includegraphics[width=0.48\textwidth]{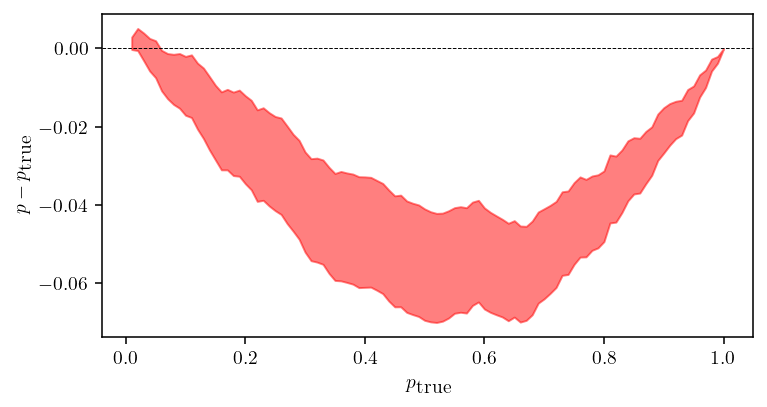}
\includegraphics[width=0.48\textwidth]{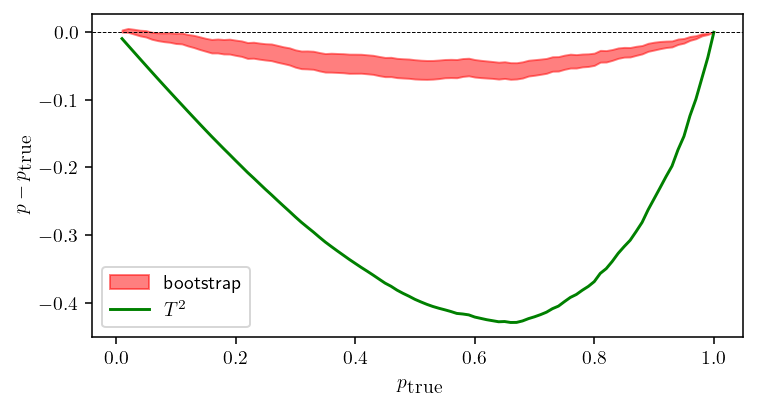}

\caption{The difference between the bootstrap and true p-values for constant fits to autocorrelated Gaussian data generated using the Metropolis algorithm with $N=800$, $T=6$, $\tau_{int} \approx 1$ and a block size of $20$. The right figure includes the corresponding difference with the $T^2$ p-value. \label{fig-autocorr-1-2} }
\end{figure}

\begin{figure}[t]
\includegraphics[width=0.48\textwidth]{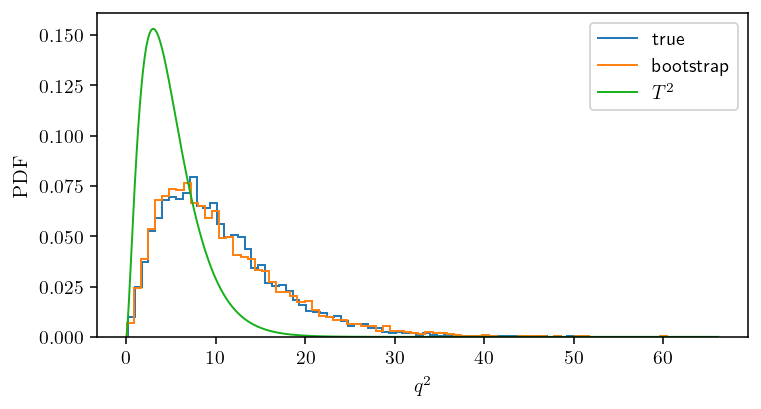}
\caption{A typical example of a bootstrap reproduction of the $q^2$ histogram for data generated as per the caption of Fig.~\ref{fig-autocorr-1-2}.\label{fig:autocorr4} }
\end{figure}

\begin{figure}[t]
\centering
\includegraphics[width=0.49\textwidth]{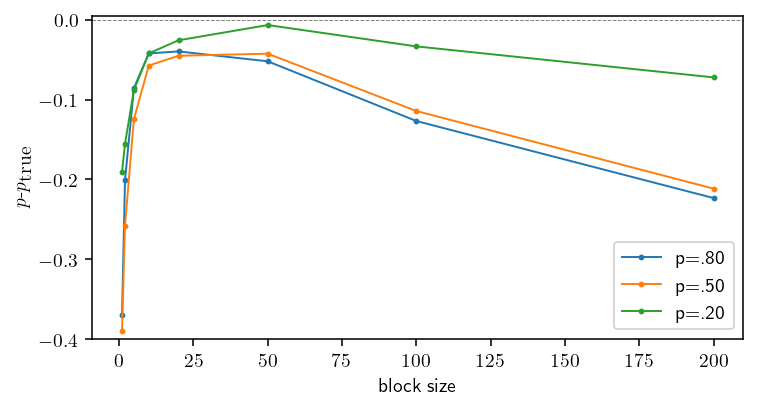}
\includegraphics[width=0.49\textwidth]{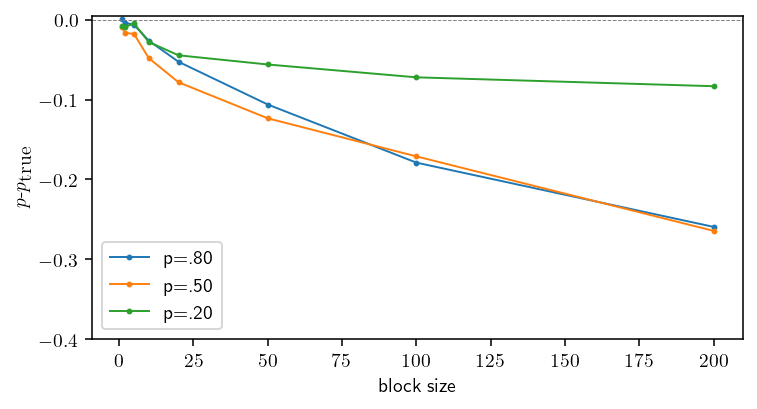}
\caption{The bias in the bootstrap p-value as a function of block size for three different choices of bootstrap p-value $p$. For each block size and value of $p$, the corresponding true p-value is obtained from our empirical true null distribution at the corresponding value of $q^2$. The left figure is for auto-correlated data generated via the Metropolis algorithm and the right figure is for independent data, both with an underlying standard normal distribution.}
\label{fig:bias_block_mcmc_and_normal}
\end{figure}

To illustrate this, we apply the procedure to autocorrelated Gaussian data with constant time dependence, generated as described in the caption of Fig.~\ref{fig-binerr-metropolis}, although here we generate 800 rather than 750 samples. We tried a number of different block sizes as shown in Fig.~\ref{fig:bias_block_mcmc_and_normal}, and found the p-value estimates plateau at around $B=20$, at which they are in agreement at the 5\% level with the true p-value (we discuss the growing deviation at large block sizes below). This value for the optimal block size is larger than the 6-8 at which the error bar plateaus in Fig.~\ref{fig-autocorr-avoid} suggesting the p-value estimate may be more sensitive to residual autocorrelation effects. In Fig.~\ref{fig-autocorr-1-2} we plot the difference between the bootstrap and true $p$-values using $B=20$, where we observe the bootstrap procedure produces an estimate accurate to a bias at the level of a few percent across the entire range of p-values -- far more reliable than the estimate one would obtain by (incorrectly) assuming the $T^2$ or $\chi^2$ distributions. In Fig.~\ref{fig:autocorr4} we compare the bootstrap, true and $T^2$ histograms of $q^2$ for a representative original ensemble, where we also see good agreement between the bootstrap and true distributions, which are vastly different in shape from the $T^2$ distribution due to the autocorrelations.

\begin{figure}[t]
\centering
\includegraphics[width=0.49\textwidth]{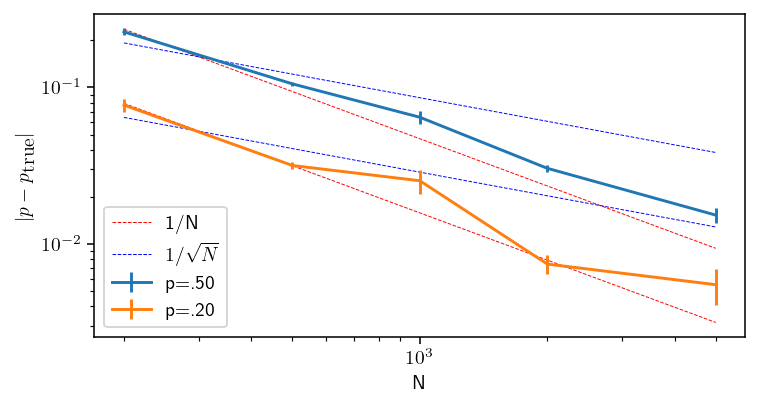}
\caption{Log-log plot of bias in $p$-value as a function of the sample size $N$ with Metropolis data. The block size is fixed at $B=50$.}
\label{fig:bias_vs_n}
\end{figure}

While the bootstrap procedure gets us much closer to an accurate $p$-value, it is of course not perfect. There are two major sources of bias. First, as we increase the block size, the number of blocks from which we are able to construct our bootstrap ensemble decreases. As discussed in Section 
\ref{sectionOrderNbias}, there is an inherent bias in the bootstrap procedure which falls off as $O(1/N_{\rm elem})$, where $N_{\rm elem}$ here is the number of individual elements in the resampling procedure. For the NBB procedure, the number of blocks, $N_{\rm elem}=N/B$, and therefore we expect the bias to grow as $B/N$, in direct proportion to the block size and inversely proportional to the ensemble size. This source of bias is responsible for the increasing deviation between the true and bootstrap p-values at large block sizes seen in Fig.~\ref{fig:bias_block_mcmc_and_normal}, which demonstrates the expected linear dependence. As this bias arises purely from the diminishing number of blocks and is not related to the autocorrelations, we observe nearly identical behavior when applying the NBB procedure to data which is not autocorrelated, as we also show in the same figure. In Fig.~\ref{fig:bias_vs_n} we also plot the bias as a function of $N$ for a fixed block size of $B=50$ (large enough to adequately account for the autocorrelations), along with lines proportional to $1/N$ and $1/\sqrt{N}$ for comparison. We observe that the bias falls convincingly like $1/N$, as expected, although this behavior is somewhat obscured by the lack of smoothness in the histograms of the true and bootstrap null distributions resulting from using a finite (but large) number of ensembles, which has a larger effect on the tails of the histogram, i.e. the low p-value regime.

In Fig.~\ref{fig:bias_block_mcmc_and_normal} we observe the large-block-size bias results in underestimated p-values, which can be explained as follows: In the limit of large block size, $B=N$, it is clear that all fluctuations in $q^2$ between resampled ensembles will cease and the bootstrap null distribution will have the form of a delta function. Given that all resampled ensembles in this limit are identical to the original, we have $\overline{C}^{*,b}(t)=\overline{C}(t)$ and thus $\overline{\epsilon}^{*,b}=0$ for all $b$ (cf. Eq.~\ref{eq-bootstrpdev}), which implies the recentered means $\bbar{\widetilde C}^{*,b}$  (cf. Eq.~\ref{eq-recentered-ens-mean}) are identical to the fitted function on the original ensemble. The $q^2$ obtained from fitting to each recentered, resampled ensemble will therefore evaluate to zero and hence the delta function describing the bootstrap null distribution will lie at the origin. The corresponding p-value will therefore evaluate to zero for any $q^2>0$. As we increase $B$ towards this limit we therefore expect the bootstrap distribution to narrow and shift towards the origin, resulting in the p-values at all values of $q^2$ evaluating to a lower values than they should have.

The second source of bias occurs as we decrease the block size, resulting in the bootstrap ensembles capturing less of the autocorrelations present in the data. As we see in Fig.~\ref{fig:bias_block_mcmc_and_normal}, this also results in p-value estimates that are lower than the true values. The reason for this can be gleaned by examining Fig.~\ref{fig:autocorr4}, where the $T^2$ curve reflects the null distribution obtained if the autocorrelations are completely ignored. We observe the true distribution is skewed to the right with a much longer tail, resulting in larger p-values for a given $q^2$. This occurs because the autocorrelations reduce the effective number of samples, which tends to broaden the null distribution as we saw in Fig.~\ref{fig:chi2vst2}. It is therefore reasonable to assume that too-small block sizes will also generally result in underestimates of the p-value.

In practice we seek a window in which the block size is large enough to capture the autocorrelations of the original ensemble while not being so large as to incur significant finite-sample bias. In our case we observe that the p-value estimate is nearly constant for block sizes in the range 20-50. However, the fact that the bias never evaluates to zero indicates that the effects of the large-block-size bias are still being felt in this regime. Assuming that such a plateau is found for a given analysis (which will depend on details of the ensemble and the fitting procedure) there is, unfortunately, no clear way to evaluate the residual bias. Nevertheless, we argue above that the p-value will be underestimated in general, which is reassuring in the context of a hypothesis test.

\section{Conclusions}

In this document we have introduced a novel technique for estimating the goodness-of-fit whereby the bootstrap method is employed to generate an approximation to the null distribution describing the fluctuations of the ``loss" (aka. $\chi^2$, or $q^2$ in this document) of the fit over many independent ensembles, from which a p-value can be directly obtained. This data-driven approach remains applicable even beyond the regimes of correlated fits with large ensemble sizes or to normal data, where analytic descriptions of the null distribution -- the $\chi^2$ and Hotelling $T^2$ distributions, respectively -- exist, allowing for arbitrary underlying data distributions and manipulations of the covariance matrix.

We demonstrated this approach in the context of a lattice quantum chromodynamics (QCD) calculation using toy data for which the true null distribution can be estimated by fitting many independent ensembles, with which we verified the applicability of the approach to both correlated and uncorrelated fits, and also to two different approaches that truncate the eigenspectrum of the covariance matrix to improve fit stability. As this approach employs a fit to the original data as part of the procedure, as part of this discussion we introduced and tested as bootstrap-based strategy for estimating the variation of our estimated p-value over different original ensembles.

An important application of the goodness-of-fit is in the context of a hypothesis test of whether the fit function is a good description of the data, where a low p-value is interpreted as indicating the presence of systematic effects in the data not accounted for by the fit function. We described how our bootstrap approach reliably estimates the null distribution for the class of hypotheses where the description of the fluctuations of the data (but not the time dependence of the mean values) is based on the data itself. While this approach is unable to account for any changes in the structure of the covariance matrix that might be associated with different hypotheses for the time dependence of the data, we nevertheless discuss the extent to which it continues to provide a reliable hypothesis test for discriminating between different fit functions, and demonstrate this using toy data.

Another important consideration for a lattice calculation is the presence of autocorrelations in the data resulting from the underlying Markov chain generation process. These effects can be accounted for by binning the data -- averaging over blocks of consecutive samples -- prior to fitting, which restores the applicability of the analytic descriptions of the null distribution (at least for correlated fits), at the cost of reducing the number of values available with which to estimate the covariance matrix, which can lead to large statistical errors on the fit parameters. We discuss an alternative ``autocorrelation-ignoring" approach employed in Ref.~\cite{Kelly:2019wfj} whereby the covariance matrix is estimated from the full, unbinned data set and the non-overlapping block bootstrap (NBB) used to estimate the standard error. In this document we demonstrate using toy data the applicability of our bootstrap method (also using NBB) for estimating the goodness-of-fit in this context, and discuss the bias incurred by using too small or too large values of the block size.

We conclude that this approach offers significant value, allowing the lattice QCD practitioner considerable freedom to modify the fitting procedure while retaining the ability to estimate a goodness-of-fit metric appropriate for a hypothesis test or otherwise.

\bibliographystyle{apsrev}
\bibliography{ref}

\end{document}